\documentclass [
 reprint,
 amsmath,amssymb,
 aps,
prb,
floatfix,
]{revtex4-2}

\usepackage{graphicx}% Include figure files
\usepackage{dcolumn}% Align table columns on decimal point
\usepackage{bm}% bold math
\usepackage{nccmath}
\usepackage{circuitikz}
\usepackage{tikz}
\usepackage{float}
\usetikzlibrary{calc,decorations.markings}
\usepackage{tcolorbox}
\usepackage[T1]{fontenc}

\begin{document}

%\preprint{APS/123-QED}

\title{Loss vs Magnetization Threshold Phenomenon for Lorentz Nonreciprocity Induced by a  Gyrotropic Particle Inside a Cavity}
\thanks{Copyright 2024 by the authors. This article is distributed under a Creative Commons Attribution 4.0 (CC BY) License}
\thanks{This material has been submitted to Physical Review Letter.}

\author{Koffi-Emmanuel Sadzi}
\email{sadzi@mail.tau.ac.il}
\author{Yakir Hadad}
\email{hadady@eng.tau.ac.il}
\affiliation{School of Electrical Engineering, Tel Aviv University, Israel, 69978}
\begin{abstract}
When a plasmonic particle is subject to a static magnetic field, $\bm{B}_{\rm dc}=B_{0} \hat{z}$, its gyrotropic response gives rise to nonreciprocal dynamics of the entire ambient surroundings.  This dynamics depends on the particle's excitation which in turn depends on the gyrotropic material damping rate $\Gamma$. Thus intuitively speaking, the heavier the gyrotropic material loss, the weaker the non-reciprocal response. This is indeed the case when the particle is located in free space. In this letter, we quantify nonreciprocity using the measure defined in Eq.~(\ref{R_coef}) and show that when the gyrotropic particle is placed inside a cavity, the nonreciprocity measure $\cal{R}$ is robust against material loss up to a certain loss threshold, $\Gamma_{th}$ that depends on the magnetic biasing $B_0$.

\end{abstract}
\maketitle

\emph{Introduction}.\textemdash Nonreciprocity has attracted the continued interests of both the physics and engineering communities over the last several decades\cite{ferrites_Lax_1962, sounas2018broadband, Lawrence_NR_flat_optics2018}. Whether linear or nonlinear, nonreciprocity is based on breaking the time-reversal symmetry. In the context of linear nonreciprocal media and for practical purposes, the time-reversal symmetry is broken using a time-reversal odd bias field such as velocity \cite{Ohad_oneway2022,godin1997reciprocity}, electric current \cite{Silveirinha_nonlocal2020}, or magnetic field \cite{Electromagnetic_Nonreciprocity_2018}. However, magnetic field biasing has predominantly been the resourceful tool \cite{pozar2011microwave} combined with anisotropic materials such as ferrites \cite{Caloz_Kodera_NR_radome,Carignan_Ferromagnetic_Nanowire_Metamaterial_2011,microwave_ferrite_device_Rodrigue_1988,ferrites_Lax_1962,zvezdin1997modern,landau1984electrodynamics,jackson1998classical}, plasma\cite{ishimaru1991electromagnetic,stix1992waves}, and two-dimensional electron-gas like surfaces such as dopped semiconductors and graphene \cite{Souma_caloz_graphene2012,giantFR_crassee2011}.
Unfortunately, the magneto-optical effect is typically weak. As a result, achieving significant nonreciprocity for practical applications often requires a bulky setup and strong magnetic biasing. Along with many works towards magnet-less solutions, several approaches to enhancing the nonreciprocal response within magneto-optical materials stand. These expand from using 1D periodic structures of microparticles \cite{Chiral_Topological_Tasolamprou2021,hadad2010magnetized,katsantonis2023giant,Du_Mori_enhanced_2010} to 2D and 3D metamaterials \cite{Fan_Nasir_NR2019,optical_isolator_Chen_Alu_2021}.

As the essence of the non-reciprocity is light-matter interactions, tailoring the underlying electromagnetic fields offers a control on interactions \cite{Reduced_dimension_Boddeti2024}. Embedding the system of interest inside an electromagnetic environment is therefore a promising route to such enhancement. In fact, coupling a system matter to a cavity is currently attractive for its potential to offer interesting consequences, among which the control of electronic phases such as superconductivity and ferro-electricity\cite{Ashida2020,Sentef2018,Curtis2019,Schlawin2019,Latini2021}, topological or magnetic phases and quantum spin liquids \cite{Mivehvar2017,Dmytruk2022,Appugliese2022,Bacciconi2023,Chiocchetta2021,Mercurio2023}.

\begin{figure}
    \centering \includegraphics[width=1.3\linewidth]{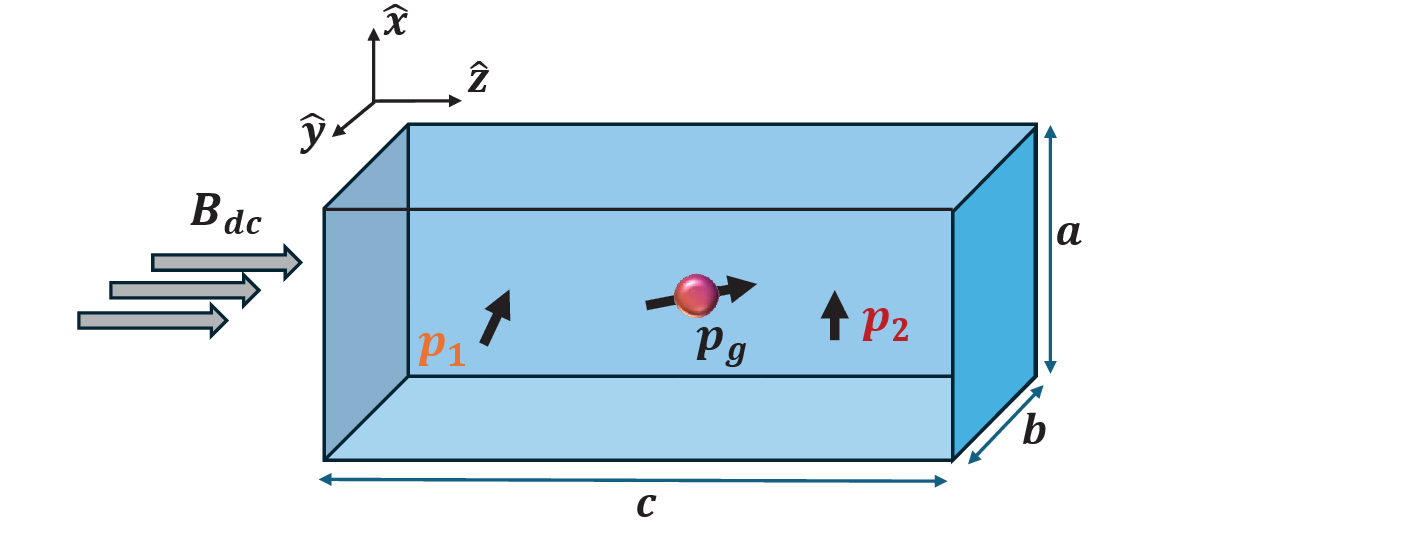}
    \caption{The physical setup: (a) A gyrotropic particle located inside a closed cavity and subject to static magnetic biasing $\bm{B}_{\rm dc}=B_0\hat{z}$. The two testing dipole sources $\bm{p}_1$ and $\bm{p}_2$ are used to calculate the nonreciprocity measure $\cal{R}$ as defined in Eq.~(\ref{R_coef}).}
    \label{setup}
\end{figure}

Here, we explore a system of a gyrotropic particle embedded in an electromagnetic cavity (Fig.~\ref{setup}). The particle is spherical, located at $\bm{r}'=(x',y',z')$, and biased with a static magnetic field $\bm{B}_{\rm dc}$.  The particle's parameters are tuned such that it resonates near the first cavity resonance, allowing for strong coupling between the two resonant systems.
%
%In a previous paper\cite{}, we showed that the coupling between the two resonators breaks the degeneracy in both the cavity and particle resonances. We now present the potential of this coupled system as a non-reciprocal 2-port device.
%
%
In this setup, we show, using numerical simulations based on the discrete dipole approximation and polarizability theory, with the exact Green's function in the cavity, that the non-reciprocity achieved in this setup exhibits peculiar robustness to loss. Specifically, we show that the nonreciprocity of the system is practically unaffected by the material loss in the particle below a critical threshold collision rate value or, equivalently, above a certain cyclotron rate threshold. This intriguing correspondence between the collision rate and the cyclotron rate, in a threshold phenomenon, is the key result of this letter.

\emph{Formulation}.\textemdash We assume that the rectangular cavity is made of ideally electric conducting walls, with dimensions given by $a=b=10\mu$m, $c=30\mu$m (thus, the first cavity resonant frequency is $ \sim 15\mbox{THz}$). The cavity's interior is characterized as a vacuum.
The radius of the gyrotropic sphere is $R_0=1\mu$m and is magnetically biased with $\bm{B}_{\rm dc}=B_0 \hat{z}$ which produces a cyclotron frequency $\omega_c=-q_e B_0/m_e$, where $q_e,m_e$ denote the elementary charge and mass of an electron, respectively. We denote by $\omega_p$ the plasma frequency and by $\Gamma$ the collision frequency of the plasmonic material of which the particle consists \cite{ishimaru1991electromagnetic}. Under these conditions, since the particle is small on the wavelength at the cavity's first resonance ($2R_0\ll 20\mu$m) and on the dimensions of the cavity ($2R_0\ll a,b,c$), it can be modeled using its electric polarizability function $\underline{\underline{\alpha}}_{e}$ that is given in \cite{SupplementaryMaterial} (Eq.~(S1)). In the absence of magnetization and when isolated in free space, this particle resonates about $\omega_p/\sqrt{3}$. In the presence of weak static magnetization ($\omega_c\ll\omega_p$), this resonance is split into three adjacent resonance frequencies, representing the lift of the modal degeneracy of the isotropic spherical particle. The cyclotron resonances that are located about $\omega_c$, much lower than the plasmonic resonance, are not considered in this work.

We wish to quantify the non-reciprocal response in the system. To that end, we place two radiating (testing) dipoles $\bm{p}_1$ and $\bm{p}_2$, at $\bm{r}_1$ and $\bm{r}_2$, respectively. To investigate the Lorentz reciprocity or rather the non-reciprocity in the system, we define the reciprocity measure $\mathcal{R}$,
\begin{align}
\label{R_coef}
    \mathcal{R}=\frac{\bm{p}_1 \cdot \bm{E}_2- \bm{p}_2 \cdot \bm{E}_1 }{|\bm{p}_1 \cdot \bm{E}_2|+ |\bm{p}_2 \cdot \bm{E}_1|}.
\end{align}
Derivations and relevant properties of  $\mathcal{R}$ in the rectangular cavity are discussed in the Supplemental Material \cite{SupplementaryMaterial}. (see also references \cite{koffi2024,novotny2012principles,collin1991field} therein).
Here $\bm{E}_1$ ($\bm{E}_2$) denotes the field at $\bm{r}_2$  ($\bm{r}_1$) due to the test dipole source $\bm{p}_1$  ($\bm{p}_2$) in the absence of $\bm{p}_2$ ($\bm{p}_1$).  With this notation,  $\bm{p}_i\cdot \bm{E}_j$ ($i\neq j$) denotes the \textit{reaction} \cite{rumsey1954reaction,kong1972theorems} between the source $i$ due to the source $j$. Given a media where Lorentz reciprocity holds, $\mathcal{R}\equiv 0$. On the other hand, from triangle inequality, one can show that $|\mathcal{R}|$ is bounded from above by 1. Thus, $0 \leq |\mathcal{R}| \leq 1$ where the lower bound corresponds to a completely reciprocal scenario, while the upper bound represents maximal nonreciprocity, as defined by this measure.

%We investigate how $\mathcal{R}$ changes as a function of frequencies. Therefore, $exp(j\omega t)$ is assumed and omitted.

\begin{figure}
\centering
\hspace*{-2cm}
\includegraphics[width=1.5\linewidth]{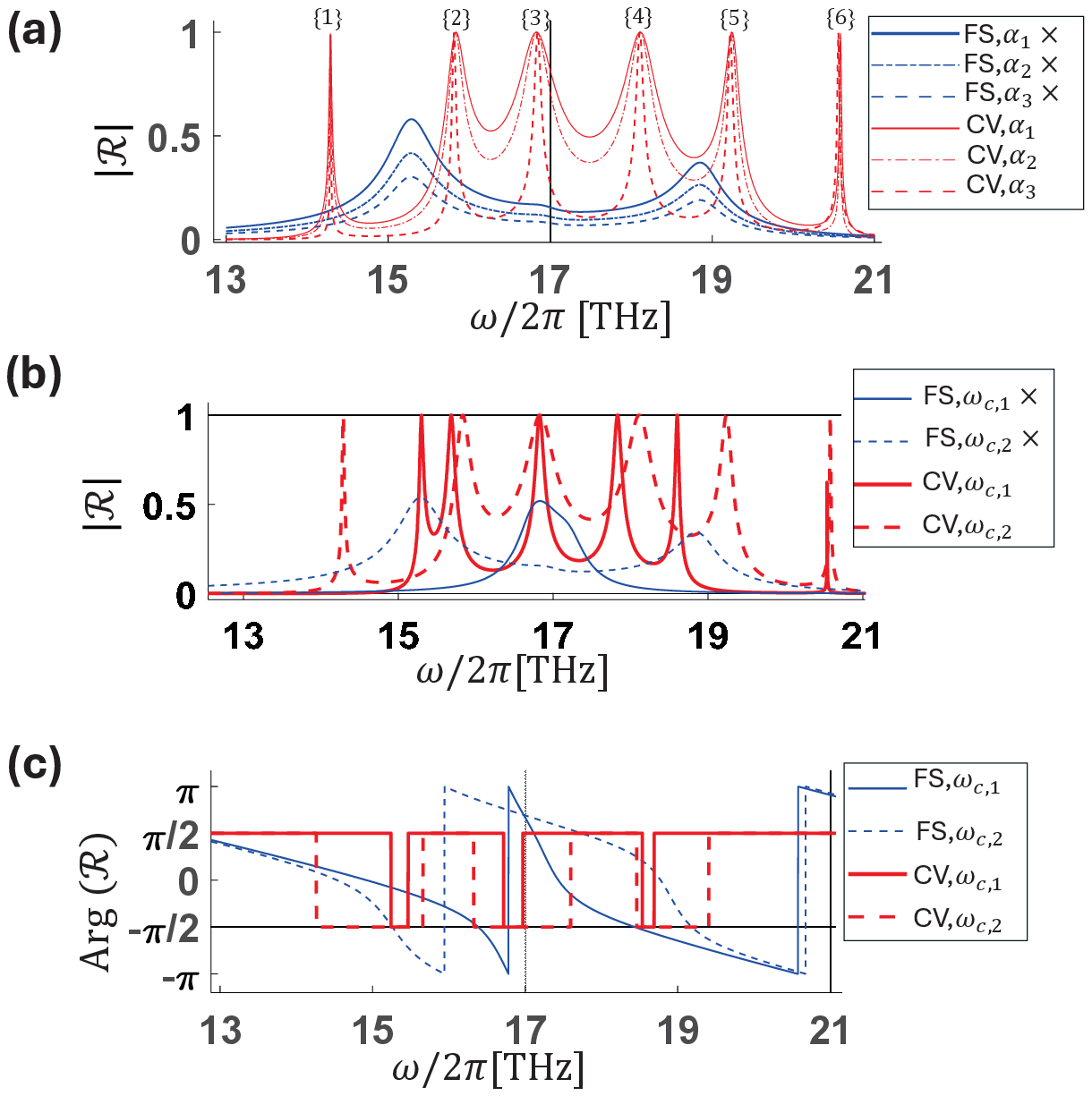}
%  \\
% \hspace{-1cm}
% \includegraphics[width=0.95\linewidth]{R_abs_eps.eps}
\caption{The reciprocity measure as a function of the excitation frequency. The notation $\times$ in the legend means that the shown curve is magnified by 100 (a) Shows the influence of the polarization of one of the testing dipoles on the reciprocity measure $\mathcal{R}$. Specifically, $\bm{p}_1=(\sin\alpha_l,\cos\alpha_l,0)$ and $\bm{p}_2=(\cos\beta,\sin\beta,0)$ with $l=1,2,3$ and $\alpha_1=30^\circ,\alpha_2=45^\circ,\alpha_3=60^\circ $ and $ \beta=30^\circ$. CV and FS indicate whether the setup is embedded inside the cavity and in free space, respectively. The numbering $\{n\}$ on top of (a) represents the $n$-th NR resonance denoted as $\omega_{r,n}$. The cyclotron frequency in all these cases is $\omega_c=0.12\omega_p$.
(b) Shows magnitude of $\mathcal{R}$  as a function of frequencies for fixed testing source polarization, with $\alpha_1$ and $\beta$ that are given in (a), but for two distinct cyclotron frequencies, $\omega_{c,1}=0.018\omega_p$ and $\omega_{c,2}=0.12\omega_p$. It is observed that the resonances in $\mathcal{R}$ broaden with the increase of $\omega_c$.(c) As in (b) but for the phase of $\mathcal{R}$. In the cavity, the phase alternates between two constants $\pm\pi/2$ where the transitions occur near the collective resonance frequencies of the system. In the free space, however, the phase varies continuously apart from the jumps between $\pi$ and $-\pi$ that can be unwrapped.
}
\label{Rabs_1}
\end{figure}

\emph{Basic dynamics of $\mathcal{R}$}.\textemdash
We first wish to get some insight regarding the dynamics of the reciprocity measure $\mathcal{R}$. To that end, in Fig.~\ref{Rabs_1}(a) we explore several basic cases. First, in the absence of the cavity, namely, when the gyrotropic particle and the testing sources are located in free space, the fields $\bm{E}_1$ and $\bm{E}_2$ can be decomposed into two terms, free-space, and scattering by the gyrotropic particle. For example, for $\bm{E}_1$ in Eq.~(\ref{R_coef}) we can write
\begin{equation}\label{eq:E1 free space}
\bm{E}_1=\underline{\underline{G}}^{\rm fs}(\bm{r}_2,\bm{r}_1)\cdot\bm{p_1} + \underline{\underline{G}}^{\rm fs}(\bm{r}_2,\bm{r}_g)\cdot\underline{\underline{\alpha}}_{e}\underline{\underline{G}}^{\rm fs}(\bm{r}_g,\bm{r}_1)\bm{p_1}
\end{equation}
where $\underline{\underline{G}}^{\rm fs}(\bm{r},\bm{r}')$ denotes the dyadic electric Green's function due to a dipole source $\bm{p}$ that is located at $\bm{r}'$, where $\bm{r}$ is the observer location. A similar expression can be obtained for $\bm{E}_2$ by a simple change in the indexes $1\longleftrightarrow2$.
Obviously, due to the reciprocity of the free space, when plugged in the numerator of Eq.~(\ref{R_coef}), the first-term in Eq.~(\ref{eq:E1 free space}) cancels with its counterpart due to $\bm{E}_2$. Consequently, as opposed to the denominator in Eq.~(\ref{R_coef}), the numerator, in this case, will strongly depend on the particle's excitation, which is expected to be relatively weak in any case due to the small particle's volume \cite{SupplementaryMaterial} (Eq.~(S1a-e)). In addition, it is expected to be severely dependent on the testing dipole orientation, as well as on the collision rate $\Gamma$ of the gyrotropic material.  This is shown by the blue curves in Fig.~\ref{Rabs_1}(a) where $|\mathcal{R}|$ is plotted (magnified by $\times100$) as a function of the frequency, around the plasmonic resonance, for different testing dipole orientations.  In contrast to the free space case, when the complete setup (i.e., the particle and the testing dipoles) is inserted into the cavity, the nonreciprocity measure is greatly enhanced, as shown by the red curves in   Fig.~\ref{Rabs_1}(a). In the numerical examples in this letter we set $\bm{r}_1=(3a/5,b/3,c/9)$, $\bm{r}_2=(a/5,3b/4,c/4)$ and  $\bm{r}_g=(0.3a,0.4b,0.7c)$ and $\left(\omega_p/2\pi\right)/\sqrt{3}=17\,$THz. Inside the cavity, the non-reciprocity peaks reach unity $|\mathcal{R}|=1$ about the collective cavity-particle resonance frequencies (see \cite{koffi2024}), and moreover, these peaks are not affected by the testing dipoles orientation. We will term the frequencies at which $\mathcal{R}=1$ as the non-reciprocity resonances.

In the cavity, provided that $\bm{p}_1,\bm{p}_2$ being real vectors (which can be assumed here with no loss of generality), and assuming lossless particle ($\Gamma=0$), the reactions in Eq.~(\ref{R_coef}) are complex conjugate of each other. Thus, a closer inspection to determine the condition underlying $|\mathcal{R}|=1$, in the cavity case, attributes the resonance condition to the requirement that $\mbox{Re}\{\bm{p}_1\cdot \bm{E}_2\}=\mbox{Re}\{\bm{p}_2\cdot \bm{E}_1\} =0$, i.e., the reactions are purely imaginary. Indeed, the resonance of the cavity-particle system forces a change (abrupt in the lossless particle case and damped in the lossy case) of sign of the imaginary part $\{\bm{p}_i\cdot \bm{E}_j\}$  followed by the nullification of the real part $\{\bm{p}_i\cdot \bm{E}_j\}$, which we will qualify as \textit{pseudo-absorption} - a manifestation of the magnetization. Although the occurrence of the \textit{pseudo-absorption} is due to resonance of the collective particle-cavity system, the \textit{pseudo-absorption} takes advantage of conjugate symmetry of the reactions, imposed by the cavity. See \cite{SupplementaryMaterial}(Eqs.~(S14-S15)).
%\footnote{The $\pi$-phase shift also occurs in the free space case, but only in $\mathcal{R}$ and in the static regime.}
A similar outcome has been proven in the quantum regime where in \cite{Rodriguez2013} the authors show a change of sign of Faraday angle when absorption processes occur, and absorption is due to excitation of particles.

\begin{figure}
\centering
    \includegraphics[width=0.98\linewidth]
    {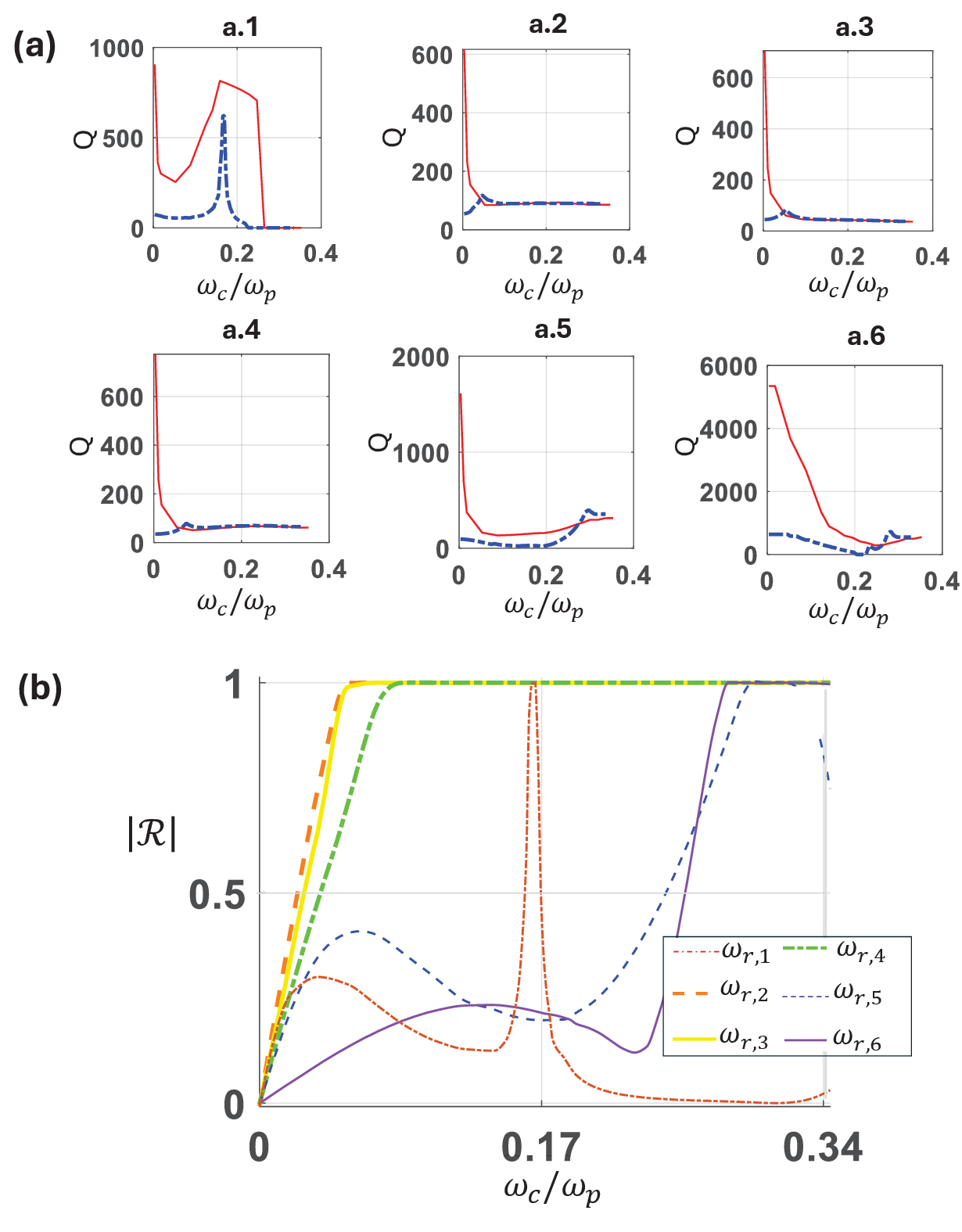}
\caption{(a) Quality factor of non-reciprocity resonances as a function of the magnetization manifested by the cyclotron frequency $\omega_c$, when the gyrotropic sphere is lossless with $\Gamma=0$ (red, solid line) and lossy with collision rate $\Gamma=0.026\omega_p$ (blue, dash-dot line). the subfigure \textbf{a.}$\bm{n}$ corresponds to
 $\omega_{r,n}$.
 (b) Shows the effect of the cyclotron frequency on the magnitude of the non-reciprocity measure $\mathcal{R}$ with $\Gamma=0.026\omega_p$. At very low magnetization, due to the loss, the nonreciprocity is negligible with $\mathcal{R}\approx0$. As the magnetization increases, $|\mathcal{R}|$ increases up to the plateau level at the maximal value for some of the resonances $n=2,3,\dots,6$ with an indicative magnetization threshold. Near $0.05\omega_p$ for $n=2,3,4$ and near $0.3\omega_p$ and $0.28\omega_p$ for $n=5,6$ respectively. For  $n=1$, $|\mathcal{R}|$ reaches in a small range of $\omega_c$ without a threshold behavior.}
\label{R_lossy_cavv}
\end{figure}

This pseudo-absorption is manifested by the finite bandwidth of the nonreciprocity resonances that are shown in Fig.~\ref{Rabs_1}. Thus, despite being calculated for a lossless problem ($\Gamma=0$), one may attribute an effective quality factor $Q$ to each one of the resonances shown in Fig.~\ref{Rabs_1}. This $Q$ factor is numerically calculated as the ratio between the resonance frequency, $\omega_n$ where $|\mathcal{R}|=1$ (in Fig.~\ref{Rabs_1}, $n=1..6$) and the bandwidth $\Delta\omega_n$ that corresponds to that resonance. The latter is simply taken as the frequency distance between the $\mathcal{R}=1/\sqrt{2}$ points around $\omega_n$.

\emph{Loss vs magnetization threshold phenomenon}.\textemdash
Up to this point, we neglected the material loss in the particle by assuming $\Gamma=0$. However, exploring the relations between the pseudo-absorption described above and the actual material loss is interesting. To that end, for each of the nonreciprocity resonances, we calculate the $Q$ factor as defined above and plot it in Fig.~\ref{R_lossy_cavv}(a) as a function of the cyclotron frequency $\omega_c$. We do that for two cases: red continuous line - assuming $\Gamma=0$ i.e., no material loss, and blue dash-dotted line - assuming lossy particle with $\Gamma=0.026\omega_p=0.8\,\mbox{THz}$

% In simple terms, the Q-factor indicates how frequency-selective is NR in our system and when do we have an optimal response.
When the system is lossless ($\Gamma=0$), the general trend is that as one increases the cyclotron frequency, there is a decrease in the $Q$ factor. Since $\Gamma=0$ in this case, the only reason for this decrease is the increase of the pseudo-absorption described above. This is an \textit{apparent} loss mechanism and in the current case, is in direct correlation with the magnetic field intensity.
In contrast, in the lossy case, with $\Gamma\neq0$, the total $Q$ factor, which comprises the material absorption and the pseudo-absorption, is, in general, significantly lower. Interestingly, the $Q$-factor in the lossy case and in the lossless case coincide above a certain magnetization threshold. Implying that in certain cases, the material loss can be neglected compared with the pseudo-absorption.  This hints at the plausibility of observing in our system a critical transition or threshold phenomenon with respect to either the loss ($\Gamma$) or the magnetization ($\omega_c$).

Such a threshold phenomenon is demonstrated in Fig.~\ref{R_lossy_cavv}(b). In the figure, we fix $\Gamma=0.026\omega_p$ and draw $|\mathcal{R}|$ as a function of $\omega_c$. Each of the curves corresponds to a different resonance, with a numbering scheme that follows that used in Fig.~\ref{Rabs_1}. Two groups of curves in Fig.~\ref{R_lossy_cavv}(b) can be distinguished. First, corresponds to resonance number 2-6, and second, corresponds to resonance number 1.
In the case of $\omega_{r,2},\omega_{r,3}$ and $\omega_{r,4}$ in Fig.\ref{R_lossy_cavv}(a.2-4) the Q-factor in the lossy case, increase and equate the lossless one above a threshold cyclotron frequency $\omega_c\sim 0.05\omega_p$. This implies that the pseudo-absorption is more dominant than the material absorption, securing robustness against the material loss mechanism. For $\omega_{r,5}$ and $\omega_{r,6}$ in Fig.\ref{R_lossy_cavv}(a.5-6), the same behavior is observed but with a much higher cyclotron frequency threshold $\omega_c\sim 0.3\omega_p$ and $\omega_c\sim 0.28\omega_p$ for n=5 and 6 respectively. In contrast, for $\omega_{r,1}$, the material loss mechanism dominates the system response for any cyclotron frequency in the range explored. Therefore, nonreciprocity measure $\mathcal{R}$ reaches 1 but could not sustained due to the strong effect of the collision rate $\Gamma$.

\begin{figure*}[t]
\includegraphics[width=0.95\textwidth]{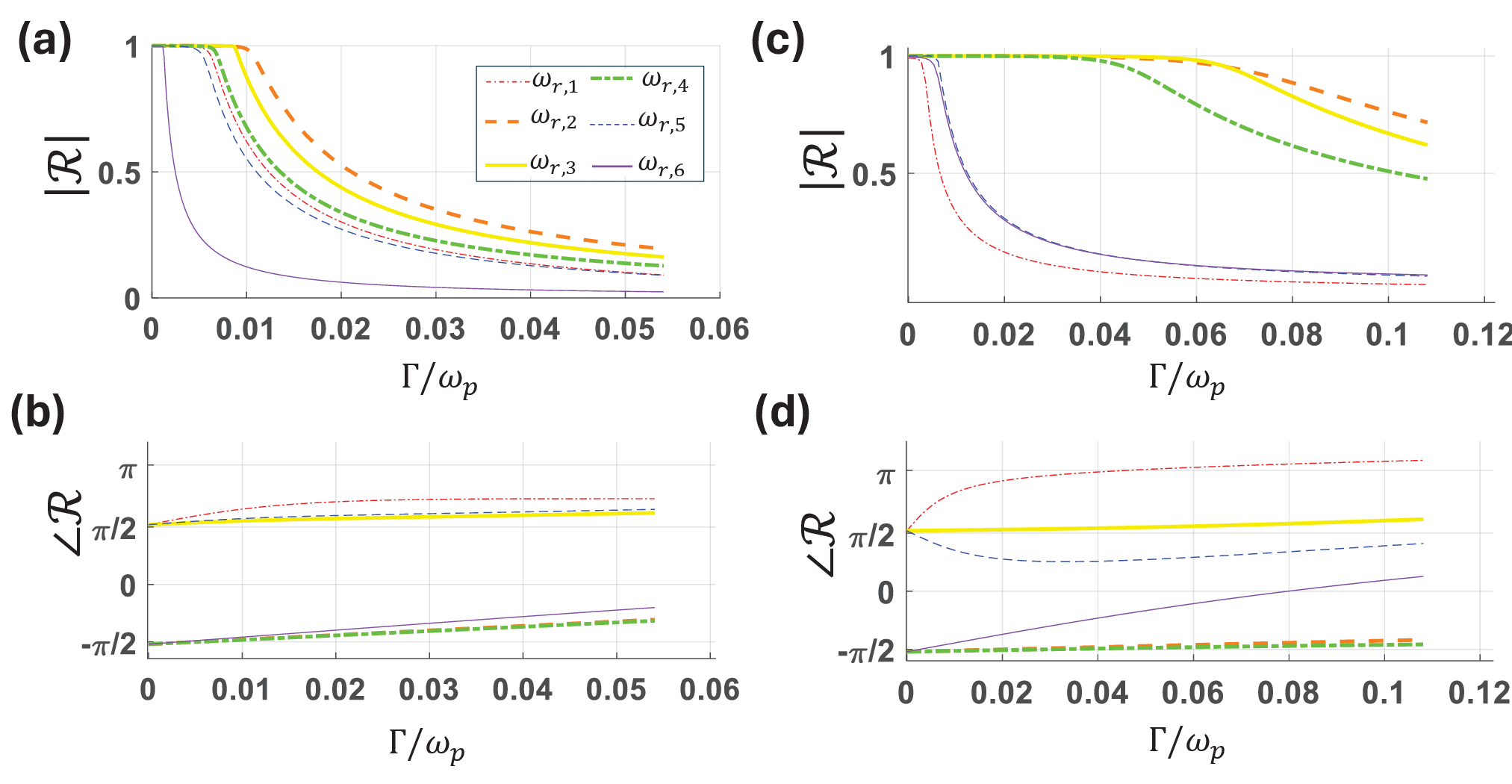}
 \caption{Loss threshold phenomenon for non-reciprocity in a collective magnetized particle-cavity system. In the absence of loss ($\Gamma=0$) for any $\omega_c\neq0$ nonreciprocity resonances, as shown in Fig.~\ref{Rabs_1}, are expected. However, as we include material loss, the magnetization value ($\omega_c$) dictates the nonreciprocity immunity to it. For a fixed magnetization, $|\mathcal{R}|$ will exhibit a plateau behavior with respect to $\Gamma$, up to a certain threshold value $\Gamma_{th,n}$ that depends on the magnetization and on the resonance number. For $\Gamma>\Gamma_{th,n}$ the nonreciprocity measure will become significantly weaker. In addition, as the $\Gamma$ increases, the phase of $\mathcal{R}$, denoted by $\angle R$, deviates from $\pm \pi/2$. A fast deviation from $\pm \pi/2$ correlates with the weakening of the nonreciprocity.
 (a-b) For $\omega_c=0.018\omega_p$ , the threshold values are $\Gamma_{th,n}<0.011\omega_p$, while for $\omega_c=0.12\omega_0$ as shown in (c-d), the loss robustness is enhanced in resonances $n=2,3,4$, as shown by the wider plateau in $|\mathcal{R}|$, and by the $\mathcal{R}$'s phase that is tightly close to $\pm \pi/2$.}
 \label{gamma_effect}
\end{figure*}

A different perspective can be obtained as we fix the static magnetization by setting a fixed value for $\omega_c$ and varying the collision rate $\Gamma$ instead.
The results in this case are shown in Fig.~\ref{gamma_effect} where we depict $\mathcal{R}$ versus $\Gamma$ for each of the six resonances shown in Fig.~\ref{Rabs_1} at $\omega_{r,n}$. For resonance number $n$, we denote by $\Gamma_{th,n}$ the largest $\Gamma$ value for which $|\mathcal{R}|$ can be considered as unity.
The trajectories of $\mathcal{R}$ as a function of the collision rate $\Gamma$ are shown in Fig.~\ref{gamma_effect}(a) for a biasing magnetic field that corresponds to $\omega_c=0.018\omega_p$. In this case, as can be seen in the six colored curves,  $\Gamma_{th,n}\le 0.011\omega_p$ for all $\omega_{r,n}$. It is also intriguing to see that the phase of $\mathcal{R}$  (Fig..~\ref{gamma_effect}(b)) deviates from $\pm \pi/2$, showing a weakening of the coupling with the cavity. Hence, the collision rate effect is significant, reducing substantially the measure $\mathcal{R}$ for all the non-reciprocity resonances.
%
%$\omega_c/{2\pi}\leq1.5\mbox{THz}$, none of the $omega_r$, especially $\omega_{r,2},\omega_{r,3}$  and  $\omega_{r,4}$ reaches $|\mathcal{R}|$, therefore the material loss strongly affects and reduces the nonreciprocity in the system for any $\Gamma$.
%
%, implying the sustaining of strong coupling to the cav,

In contrast, as we enhance the magnetization with respect to the previous case by a factor of $\sim 7$, such that $\omega_c=0.12\omega_p$, we obtain that $\mathcal{R}$ is robust against material loss for a much wider range of $\Gamma$, as shown in Fig.~\ref{gamma_effect}(c-d).
Specifically, a collision rate threshold as high as $\Gamma_{th,2}\approx\Gamma_{th,3}\approx0.065\omega_p$ for $\omega_{r,2}$, and $\omega_{r,3}$, and $\Gamma_{th,4}\approx 0.045\omega_p$  for $\omega_{r,4}$ are obtained. Likewise, the phase of $\mathcal{R}$ at $\omega_{r,2}$, $\omega_{r,2}$ and $\omega_{r,4}$ is maintained close to $\pm \pi/2$ while the remaining ones show a fast shift of the phase away from $\pm \pi/2$. This establishes the relation between the magnetization and the collision rate and in particular, demonstrates the loss threshold phenomenon in the collective magnetized particle-cavity system, which is the key result of this work.

%\begin{widetext}
%

\emph{Conclusions and discussion}.\textemdash In this letter, we explored Lorentz nonreciprocity due to a magnetized plasmonic particle inside a cavity. In particular, we defined a non-reciprocity measure $\mathcal{R}$ by Eq.~(\ref{R_coef}) and demonstrated its frequency dynamics for free-space and inside the cavity setups. In particular, we showed that nonreciprocity peaks can be obtained inside the cavity, and these correspond to the resonances of the collective particle-cavity system. Interestingly, we showed that cavity nonreciprocity can be robust against certain loss levels, where the robustness loss threshold depends on the magnetization (manifested by the cyclotron frequency). We explain this robustness by noting that through the defined non-reciprocity measure $\mathcal{R}$ one can observe a pseudo-absorption phenomenon that determines the bandwidth of the nonreciprocity resonances in $\mathcal{R}$, and thus, in turn, also their $Q$ factor. In the presence of material loss, which is described by a finite collision rate $\Gamma$ these two loss mechanisms compete. In cases where the material loss effect is much weaker than the pseudo-absorption effect, the material loss effect on the nonreciprocity will practically not be seen.
This competing loss mechanism agrees, in a different context,  with one of the main advantages of plasmonics use stated in\cite{burla2023} suggesting the operation of plasmonic modulators in the vicinity of absorption resonances of an electro-optical material. For example \cite{Haffner2017} showed that the material losses, even when amplified by resonance, contribute minimally to the overall losses of the plasmonic waveguide, and \cite{Yariv_sagnac2006,Steinberg_CROW2012}, showed a similar idea applied for photonic crystal waveguides.

%\bibliography{main}

\begin{thebibliography}{50}%
\sloppy
\makeatletter
\providecommand \@ifxundefined [1]{%
 \@ifx{#1\undefined}
}%
\providecommand \@ifnum [1]{%
 \ifnum #1\expandafter \@firstoftwo
 \else \expandafter \@secondoftwo
 \fi
}%
\providecommand \@ifx [1]{%
 \ifx #1\expandafter \@firstoftwo
 \else \expandafter \@secondoftwo
 \fi
}%
\providecommand \natexlab [1]{#1}%
\providecommand \enquote  [1]{``#1''}%
\providecommand \bibnamefont  [1]{#1}%
\providecommand \bibfnamefont [1]{#1}%
\providecommand \citenamefont [1]{#1}%
\providecommand \href@noop [0]{\@secondoftwo}%
\providecommand \href [0]{\begingroup \@sanitize@url \@href}%
\providecommand \@href[1]{\@@startlink{#1}\@@href}%
\providecommand \@@href[1]{\endgroup#1\@@endlink}%
\providecommand \@sanitize@url [0]{\catcode `\\12\catcode `\$12\catcode
  `\&12\catcode `\#12\catcode `\^12\catcode `\_12\catcode `\%12\relax}%
\providecommand \@@startlink[1]{}%
\providecommand \@@endlink[0]{}%
\providecommand \url  [0]{\begingroup\@sanitize@url \@url }%
\providecommand \@url [1]{\endgroup\@href {#1}{\urlprefix }}%
\providecommand \urlprefix  [0]{URL }%
\providecommand \Eprint [0]{\href }%
\providecommand \doibase [0]{https://doi.org/}%
\providecommand \selectlanguage [0]{\@gobble}%
\providecommand \bibinfo  [0]{\@secondoftwo}%
\providecommand \bibfield  [0]{\@secondoftwo}%
\providecommand \translation [1]{[#1]}%
\providecommand \BibitemOpen [0]{}%
\providecommand \bibitemStop [0]{}%
\providecommand \bibitemNoStop [0]{.\EOS\space}%
\providecommand \EOS [0]{\spacefactor3000\relax}%
\providecommand \BibitemShut  [1]{\csname bibitem#1\endcsname}%
\let\auto@bib@innerbib\@empty
%</preamble>
\bibitem [{\citenamefont {Lax}\ and\ \citenamefont
  {Button}(1962)}]{ferrites_Lax_1962}%
  \BibitemOpen
  \bibfield  {author} {\bibinfo {author} {\bibfnamefont {B.}~\bibnamefont
  {Lax}}\ and\ \bibinfo {author} {\bibfnamefont {K.~J.}\ \bibnamefont
  {Button}},\ }\href@noop {} {\emph {\bibinfo {title} {Microwave Ferrites and
  Ferrimagnetics}}}\ (\bibinfo  {publisher} {McGraw-Hill},\ \bibinfo {address}
  {New York},\ \bibinfo {year} {1962})\BibitemShut {NoStop}%
\bibitem [{\citenamefont {Sounas}\ \emph {et~al.}(2018)\citenamefont {Sounas},
  \citenamefont {Soric},\ and\ \citenamefont {Alù}}]{sounas2018broadband}%
  \BibitemOpen
  \bibfield  {author} {\bibinfo {author} {\bibfnamefont {D.~L.}\ \bibnamefont
  {Sounas}}, \bibinfo {author} {\bibfnamefont {J.}~\bibnamefont {Soric}},\ and\
  \bibinfo {author} {\bibfnamefont {A.}~\bibnamefont {Alù}},\ }\bibfield
  {title} {\bibinfo {title} {Broadband passive isolators based on coupled
  nonlinear resonances},\ }\href {https://doi.org/10.1038/s41928-018-0025-0}
  {\bibfield  {journal} {\bibinfo  {journal} {Nature Electronics}\ }\textbf
  {\bibinfo {volume} {1}},\ \bibinfo {pages} {113} (\bibinfo {year}
  {2018})}\BibitemShut {NoStop}%
\bibitem [{\citenamefont {Lawrence}\ \emph {et~al.}(2018)\citenamefont
  {Lawrence}, \citenamefont {Barton},\ and\ \citenamefont
  {Dionne}}]{Lawrence_NR_flat_optics2018}%
  \BibitemOpen
  \bibfield  {author} {\bibinfo {author} {\bibfnamefont {M.}~\bibnamefont
  {Lawrence}}, \bibinfo {author} {\bibfnamefont {D.~R.~I.}\ \bibnamefont
  {Barton}},\ and\ \bibinfo {author} {\bibfnamefont {J.~A.}\ \bibnamefont
  {Dionne}},\ }\bibfield  {title} {\bibinfo {title} {Nonreciprocal flat optics
  with silicon metasurfaces},\ }\href
  {https://doi.org/10.1021/acs.nanolett.7b04646} {\bibfield  {journal}
  {\bibinfo  {journal} {Nano Letters}\ }\textbf {\bibinfo {volume} {18}},\
  \bibinfo {pages} {1104} (\bibinfo {year} {2018})},\ \bibinfo {note} {pMID:
  29369641},\ \Eprint
  {https://arxiv.org/abs/https://doi.org/10.1021/acs.nanolett.7b04646}
  {https://doi.org/10.1021/acs.nanolett.7b04646} \BibitemShut {NoStop}%
\bibitem [{\citenamefont {Silbiger}\ and\ \citenamefont
  {Hadad}(2022)}]{Ohad_oneway2022}%
  \BibitemOpen
  \bibfield  {author} {\bibinfo {author} {\bibfnamefont {O.}~\bibnamefont
  {Silbiger}}\ and\ \bibinfo {author} {\bibfnamefont {Y.}~\bibnamefont
  {Hadad}},\ }\bibfield  {title} {\bibinfo {title} {One-way acoustic guiding
  under transverse fluid flow},\ }\href
  {https://doi.org/10.1103/PhysRevApplied.17.064058} {\bibfield  {journal}
  {\bibinfo  {journal} {Phys. Rev. Appl.}\ }\textbf {\bibinfo {volume} {17}},\
  \bibinfo {pages} {064058} (\bibinfo {year} {2022})}\BibitemShut {NoStop}%
\bibitem [{\citenamefont {Godin}(1997)}]{godin1997reciprocity}%
  \BibitemOpen
  \bibfield  {author} {\bibinfo {author} {\bibfnamefont {O.~A.}\ \bibnamefont
  {Godin}},\ }\bibfield  {title} {\bibinfo {title} {Reciprocity and energy
  theorems for waves in a compressible inhomogeneous moving fluid},\
  }\href@noop {} {\bibfield  {journal} {\bibinfo  {journal} {Wave Motion}\
  }\textbf {\bibinfo {volume} {25}},\ \bibinfo {pages} {143} (\bibinfo {year}
  {1997})}\BibitemShut {NoStop}%
\bibitem [{\citenamefont {Morgado}\ and\ \citenamefont
  {Silveirinha}(2020)}]{Silveirinha_nonlocal2020}%
  \BibitemOpen
  \bibfield  {author} {\bibinfo {author} {\bibfnamefont {T.~A.}\ \bibnamefont
  {Morgado}}\ and\ \bibinfo {author} {\bibfnamefont {M.~G.}\ \bibnamefont
  {Silveirinha}},\ }\bibfield  {title} {\bibinfo {title} {Nonlocal effects and
  enhanced nonreciprocity in current-driven graphene systems},\ }\href
  {https://doi.org/10.1103/PhysRevB.102.075102} {\bibfield  {journal} {\bibinfo
   {journal} {Phys. Rev. B}\ }\textbf {\bibinfo {volume} {102}},\ \bibinfo
  {pages} {075102} (\bibinfo {year} {2020})}\BibitemShut {NoStop}%
\bibitem [{\citenamefont {Caloz}\ \emph {et~al.}(2018)\citenamefont {Caloz},
  \citenamefont {Al\`u}, \citenamefont {Tretyakov}, \citenamefont {Sounas},
  \citenamefont {Achouri},\ and\ \citenamefont
  {Deck-L\'eger}}]{Electromagnetic_Nonreciprocity_2018}%
  \BibitemOpen
  \bibfield  {author} {\bibinfo {author} {\bibfnamefont {C.}~\bibnamefont
  {Caloz}}, \bibinfo {author} {\bibfnamefont {A.}~\bibnamefont {Al\`u}},
  \bibinfo {author} {\bibfnamefont {S.}~\bibnamefont {Tretyakov}}, \bibinfo
  {author} {\bibfnamefont {D.}~\bibnamefont {Sounas}}, \bibinfo {author}
  {\bibfnamefont {K.}~\bibnamefont {Achouri}},\ and\ \bibinfo {author}
  {\bibfnamefont {Z.-L.}\ \bibnamefont {Deck-L\'eger}},\ }\bibfield  {title}
  {\bibinfo {title} {Electromagnetic nonreciprocity},\ }\href
  {https://doi.org/10.1103/PhysRevApplied.10.047001} {\bibfield  {journal}
  {\bibinfo  {journal} {Phys. Rev. Appl.}\ }\textbf {\bibinfo {volume} {10}},\
  \bibinfo {pages} {047001} (\bibinfo {year} {2018})}\BibitemShut {NoStop}%
\bibitem [{\citenamefont {Pozar}(2011)}]{pozar2011microwave}%
  \BibitemOpen
  \bibfield  {author} {\bibinfo {author} {\bibfnamefont {D.~M.}\ \bibnamefont
  {Pozar}},\ }\href@noop {} {\emph {\bibinfo {title} {Microwave
  Engineering}}},\ \bibinfo {edition} {4th}\ ed.\ (\bibinfo  {publisher}
  {Wiley},\ \bibinfo {year} {2011})\ Chap.~\bibinfo {chapter} {9}\BibitemShut
  {NoStop}%
\bibitem [{\citenamefont {Parsa}\ \emph {et~al.}(2011)\citenamefont {Parsa},
  \citenamefont {Kodera},\ and\ \citenamefont
  {Caloz}}]{Caloz_Kodera_NR_radome}%
  \BibitemOpen
  \bibfield  {author} {\bibinfo {author} {\bibfnamefont {A.}~\bibnamefont
  {Parsa}}, \bibinfo {author} {\bibfnamefont {T.}~\bibnamefont {Kodera}},\ and\
  \bibinfo {author} {\bibfnamefont {C.}~\bibnamefont {Caloz}},\ }\bibfield
  {title} {\bibinfo {title} {Ferrite based non-reciprocal radome, generalized
  scattering matrix analysis and experimental demonstration},\ }\href
  {https://doi.org/10.1109/TAP.2010.2103016} {\bibfield  {journal} {\bibinfo
  {journal} {IEEE Trans. Antennas Propag.}\ }\textbf {\bibinfo
  {volume} {59}},\ \bibinfo {pages} {810} (\bibinfo {year} {2011})}\BibitemShut
  {NoStop}%
\bibitem [{\citenamefont {Carignan}\ \emph {et~al.}(2011)\citenamefont
  {Carignan}, \citenamefont {Yelon}, \citenamefont {Menard},\ and\
  \citenamefont {Caloz}}]{Carignan_Ferromagnetic_Nanowire_Metamaterial_2011}%
  \BibitemOpen
  \bibfield  {author} {\bibinfo {author} {\bibfnamefont {L.-P.}\ \bibnamefont
  {Carignan}}, \bibinfo {author} {\bibfnamefont {A.}~\bibnamefont {Yelon}},
  \bibinfo {author} {\bibfnamefont {D.}~\bibnamefont {Menard}},\ and\ \bibinfo
  {author} {\bibfnamefont {C.}~\bibnamefont {Caloz}},\ }\bibfield  {title}
  {\bibinfo {title} {Ferromagnetic nanowire metamaterials: Theory and
  applications},\ }\href {https://doi.org/10.1109/TMTT.2011.2163202} {\bibfield
   {journal} {\bibinfo  {journal} {IEEE Trans. Microw. Theory Tech.}\ }\textbf {\bibinfo {volume} {59}},\ \bibinfo {pages} {2568}
  (\bibinfo {year} {2011})}\BibitemShut {NoStop}%
\bibitem [{\citenamefont
  {Rodrigue}(1988)}]{microwave_ferrite_device_Rodrigue_1988}%
  \BibitemOpen
  \bibfield  {author} {\bibinfo {author} {\bibfnamefont {G.}~\bibnamefont
  {Rodrigue}},\ }\bibfield  {title} {\bibinfo {title} {A generation of
  microwave ferrite devices},\ }\href {https://doi.org/10.1109/5.4389}
  {\bibfield  {journal} {\bibinfo  {journal} {Proceedings of the IEEE}\
  }\textbf {\bibinfo {volume} {76}},\ \bibinfo {pages} {121} (\bibinfo {year}
  {1988})}\BibitemShut {NoStop}%
\bibitem [{\citenamefont {Zvezdin}\ and\ \citenamefont
  {Kotov}(1997)}]{zvezdin1997modern}%
  \BibitemOpen
  \bibfield  {author} {\bibinfo {author} {\bibfnamefont {A.}~\bibnamefont
  {Zvezdin}}\ and\ \bibinfo {author} {\bibfnamefont {V.}~\bibnamefont
  {Kotov}},\ }\href {https://doi.org/10.1201/9780367802608} {\emph {\bibinfo
  {title} {Modern Magnetooptics and Magnetooptical Materials}}},\ \bibinfo
  {edition} {1st}\ ed.\ (\bibinfo  {publisher} {CRC Press},\ \bibinfo {year}
  {1997})\BibitemShut {NoStop}%
\bibitem [{\citenamefont {Landau}\ \emph {et~al.}(1984)\citenamefont {Landau},
  \citenamefont {Pitaevskii},\ and\ \citenamefont
  {Lifshitz}}]{landau1984electrodynamics}%
  \BibitemOpen
  \bibfield  {author} {\bibinfo {author} {\bibfnamefont {L.~D.}\ \bibnamefont
  {Landau}}, \bibinfo {author} {\bibfnamefont {L.~P.}\ \bibnamefont
  {Pitaevskii}},\ and\ \bibinfo {author} {\bibfnamefont {E.~M.}\ \bibnamefont
  {Lifshitz}},\ }\href@noop {} {\emph {\bibinfo {title} {Electrodynamics of
  Continuous Media}}},\ \bibinfo {edition} {2nd}\ ed.,\ \bibinfo {series}
  {Course of Theoretical Physics}, Vol.~\bibinfo {volume} {8}\ (\bibinfo
  {publisher} {Butterworth-Heinemann},\ \bibinfo {year} {1984})\ Chap.~\bibinfo
  {chapter} {12}\BibitemShut {NoStop}%
\bibitem [{\citenamefont {Jackson}(1998)}]{jackson1998classical}%
  \BibitemOpen
  \bibfield  {author} {\bibinfo {author} {\bibfnamefont {J.~D.}\ \bibnamefont
  {Jackson}},\ }\href@noop {} {\emph {\bibinfo {title} {Classical
  Electrodynamics}}},\ \bibinfo {edition} {3rd}\ ed.\ (\bibinfo  {publisher}
  {Wiley},\ \bibinfo {year} {1998})\BibitemShut {NoStop}%
\bibitem [{\citenamefont {Ishimaru}(1991)}]{ishimaru1991electromagnetic}%
  \BibitemOpen
  \bibfield  {author} {\bibinfo {author} {\bibfnamefont {A.}~\bibnamefont
  {Ishimaru}},\ }\href@noop {} {\emph {\bibinfo {title} {Electromagnetic wave
  propagation, radiation, and scattering}}}\ (\bibinfo  {publisher} {Prentice
  Hall},\ \bibinfo {address} {Englewood Cliffs, NJ},\ \bibinfo {year} {1991})\
  pp.\ \bibinfo {pages} {xviii, 637},\ \bibinfo {note} {includes
  bibliographical references (pages 626-632) and index}\BibitemShut {NoStop}%
\bibitem [{\citenamefont {Stix}(1992)}]{stix1992waves}%
  \BibitemOpen
  \bibfield  {author} {\bibinfo {author} {\bibfnamefont {T.~H.}\ \bibnamefont
  {Stix}},\ }\href@noop {} {\emph {\bibinfo {title} {Waves in Plasmas}}}\
  (\bibinfo  {publisher} {American Institute of Physics},\ \bibinfo {year}
  {1992})\BibitemShut {NoStop}%
\bibitem [{\citenamefont {Sounas}\ and\ \citenamefont
  {Caloz}(2012)}]{Souma_caloz_graphene2012}%
  \BibitemOpen
  \bibfield  {author} {\bibinfo {author} {\bibfnamefont {D.~L.}\ \bibnamefont
  {Sounas}}\ and\ \bibinfo {author} {\bibfnamefont {C.}~\bibnamefont {Caloz}},\
  }\bibfield  {title} {\bibinfo {title} {Gyrotropy and nonreciprocity of
  graphene for microwave applications},\ }\href
  {https://doi.org/10.1109/TMTT.2011.2182205} {\bibfield  {journal} {\bibinfo
  {journal} {IEEE Trans. Microw. Theory Tech.}\ }\textbf
  {\bibinfo {volume} {60}},\ \bibinfo {pages} {901} (\bibinfo {year}
  {2012})}\BibitemShut {NoStop}%
\bibitem [{\citenamefont {Crassee}\ \emph {et~al.}(2011)\citenamefont
  {Crassee}, \citenamefont {Levallois}, \citenamefont {Walter}, \citenamefont
  {van~der Marel}, \citenamefont {Schwartz}, \citenamefont {Gallais},
  \citenamefont {Sac{\'e}p{\'e}}, \citenamefont {de~Heer}, \citenamefont
  {Berger},\ and\ \citenamefont {Kuzmenko}}]{giantFR_crassee2011}%
  \BibitemOpen
  \bibfield  {author} {\bibinfo {author} {\bibfnamefont {I.}~\bibnamefont
  {Crassee}}, \bibinfo {author} {\bibfnamefont {J.}~\bibnamefont {Levallois}},
  \bibinfo {author} {\bibfnamefont {A.}~\bibnamefont {Walter}}, \bibinfo
  {author} {\bibfnamefont {D.}~\bibnamefont {van~der Marel}}, \bibinfo {author}
  {\bibfnamefont {T.}~\bibnamefont {Schwartz}}, \bibinfo {author}
  {\bibfnamefont {Y.}~\bibnamefont {Gallais}}, \bibinfo {author} {\bibfnamefont
  {B.}~\bibnamefont {Sac{\'e}p{\'e}}}, \bibinfo {author} {\bibfnamefont
  {W.~A.}\ \bibnamefont {de~Heer}}, \bibinfo {author} {\bibfnamefont
  {C.}~\bibnamefont {Berger}},\ and\ \bibinfo {author} {\bibfnamefont {A.~B.}\
  \bibnamefont {Kuzmenko}},\ }\bibfield  {title} {\bibinfo {title} {Giant
  faraday rotation in single- and multilayer graphene},\ }\href
  {https://doi.org/10.1038/nphys1816} {\bibfield  {journal} {\bibinfo
  {journal} {Nature Physics}\ }\textbf {\bibinfo {volume} {7}},\ \bibinfo
  {pages} {48} (\bibinfo {year} {2011})}\BibitemShut {NoStop}%
\bibitem [{\citenamefont {Tasolamprou}\ \emph {et~al.}(2021)\citenamefont
  {Tasolamprou}, \citenamefont {Kafesaki}, \citenamefont {Soukoulis},
  \citenamefont {Economou},\ and\ \citenamefont
  {Koschny}}]{Chiral_Topological_Tasolamprou2021}%
  \BibitemOpen
  \bibfield  {author} {\bibinfo {author} {\bibfnamefont {A.~C.}\ \bibnamefont
  {Tasolamprou}}, \bibinfo {author} {\bibfnamefont {M.}~\bibnamefont
  {Kafesaki}}, \bibinfo {author} {\bibfnamefont {C.~M.}\ \bibnamefont
  {Soukoulis}}, \bibinfo {author} {\bibfnamefont {E.~N.}\ \bibnamefont
  {Economou}},\ and\ \bibinfo {author} {\bibfnamefont {T.}~\bibnamefont
  {Koschny}},\ }\bibfield  {title} {\bibinfo {title} {Chiral topological
  surface states on a finite square photonic crystal bounded by air},\ }\href
  {https://doi.org/10.1103/PhysRevApplied.16.044011} {\bibfield  {journal}
  {\bibinfo  {journal} {Phys. Rev. Appl.}\ }\textbf {\bibinfo {volume} {16}},\
  \bibinfo {pages} {044011} (\bibinfo {year} {2021})}\BibitemShut {NoStop}%
\bibitem [{\citenamefont {Hadad}\ and\ \citenamefont
  {Steinberg}(2010)}]{hadad2010magnetized}%
  \BibitemOpen
  \bibfield  {author} {\bibinfo {author} {\bibfnamefont {Y.}~\bibnamefont
  {Hadad}}\ and\ \bibinfo {author} {\bibfnamefont {B.~Z.}\ \bibnamefont
  {Steinberg}},\ }\bibfield  {title} {\bibinfo {title} {Magnetized spiral
  chains of plasmonic ellipsoids for one-way optical waveguides},\ }\href
  {https://doi.org/10.1103/PhysRevLett.105.233904} {\bibfield  {journal}
  {\bibinfo  {journal} {Physical Review Letters}\ }\textbf {\bibinfo {volume}
  {105}},\ \bibinfo {pages} {233904} (\bibinfo {year} {2010})}\BibitemShut
  {NoStop}%
\bibitem [{\citenamefont {Katsantonis}\ \emph {et~al.}(2023)\citenamefont
  {Katsantonis}, \citenamefont {Tasolamprou}, \citenamefont {Koschny},
  \citenamefont {Soukoulis},\ and\ \citenamefont
  {Kafesaki}}]{katsantonis2023giant}%
  \BibitemOpen
  \bibfield  {author} {\bibinfo {author} {\bibfnamefont {I.}~\bibnamefont
  {Katsantonis}}, \bibinfo {author} {\bibfnamefont {A.~C.}\ \bibnamefont
  {Tasolamprou}}, \bibinfo {author} {\bibfnamefont {T.}~\bibnamefont
  {Koschny}}, \bibinfo {author} {\bibfnamefont {C.~M.}\ \bibnamefont
  {Soukoulis}},\ and\ \bibinfo {author} {\bibfnamefont {M.}~\bibnamefont
  {Kafesaki}},\ }\bibfield  {title} {\bibinfo {title} {Giant enhancement of
  nonreciprocity in gyrotropic heterostructures},\ }\href
  {https://doi.org/10.1038/s41598-023-48503-9} {\bibfield  {journal} {\bibinfo
  {journal} {Scientific Reports}\ }\textbf {\bibinfo {volume} {13}},\ \bibinfo
  {pages} {21986} (\bibinfo {year} {2023})},\ \bibinfo {note} {received 15
  October 2023, Accepted 27 November 2023, Published 11 December
  2023}\BibitemShut {NoStop}%
\bibitem [{\citenamefont {Du}\ \emph {et~al.}(2010)\citenamefont {Du},
  \citenamefont {Mori}, \citenamefont {Suzuki}, \citenamefont {Saito},
  \citenamefont {Fukuda},\ and\ \citenamefont
  {Takahashi}}]{Du_Mori_enhanced_2010}%
  \BibitemOpen
  \bibfield  {author} {\bibinfo {author} {\bibfnamefont {G.~X.}\ \bibnamefont
  {Du}}, \bibinfo {author} {\bibfnamefont {T.}~\bibnamefont {Mori}}, \bibinfo
  {author} {\bibfnamefont {M.}~\bibnamefont {Suzuki}}, \bibinfo {author}
  {\bibfnamefont {S.}~\bibnamefont {Saito}}, \bibinfo {author} {\bibfnamefont
  {H.}~\bibnamefont {Fukuda}},\ and\ \bibinfo {author} {\bibfnamefont
  {M.}~\bibnamefont {Takahashi}},\ }\bibfield  {title} {\bibinfo {title}
  {{Evidence of localized surface plasmon enhanced magneto-optical effect in
  nanodisk array}},\ }\href {https://doi.org/10.1063/1.3334726} {\bibfield
  {journal} {\bibinfo  {journal} {Applied Physics Letters}\ }\textbf {\bibinfo
  {volume} {96}},\ \bibinfo {pages} {081915} (\bibinfo {year}
  {2010})}\BibitemShut {NoStop}%
\bibitem [{\citenamefont {Fan}\ \emph {et~al.}(2019)\citenamefont {Fan},
  \citenamefont {Nasir}, \citenamefont {Nicholls}, \citenamefont {Zayats},\
  and\ \citenamefont {Podolskiy}}]{Fan_Nasir_NR2019}%
  \BibitemOpen
  \bibfield  {author} {\bibinfo {author} {\bibfnamefont {B.}~\bibnamefont
  {Fan}}, \bibinfo {author} {\bibfnamefont {M.~E.}\ \bibnamefont {Nasir}},
  \bibinfo {author} {\bibfnamefont {L.~H.}\ \bibnamefont {Nicholls}}, \bibinfo
  {author} {\bibfnamefont {A.~V.}\ \bibnamefont {Zayats}},\ and\ \bibinfo
  {author} {\bibfnamefont {V.~A.}\ \bibnamefont {Podolskiy}},\ }\bibfield
  {title} {\bibinfo {title} {Magneto-optical metamaterials: Nonreciprocal
  transmission and faraday effect enhancement},\ }\href
  {https://doi.org/https://doi.org/10.1002/adom.201801420} {\bibfield
  {journal} {\bibinfo  {journal} {Advanced Optical Materials}\ }\textbf
  {\bibinfo {volume} {7}},\ \bibinfo {pages} {1801420} (\bibinfo {year}
  {2019})}\BibitemShut {NoStop}%
\bibitem [{\citenamefont {Kawaguchi}\ \emph {et~al.}(2021)\citenamefont
  {Kawaguchi}, \citenamefont {Li}, \citenamefont {Chen}, \citenamefont {Menon},
  \citenamefont {Alù},\ and\ \citenamefont
  {Khanikaev}}]{optical_isolator_Chen_Alu_2021}%
  \BibitemOpen
  \bibfield  {author} {\bibinfo {author} {\bibfnamefont {Y.}~\bibnamefont
  {Kawaguchi}}, \bibinfo {author} {\bibfnamefont {M.}~\bibnamefont {Li}},
  \bibinfo {author} {\bibfnamefont {K.}~\bibnamefont {Chen}}, \bibinfo {author}
  {\bibfnamefont {V.}~\bibnamefont {Menon}}, \bibinfo {author} {\bibfnamefont
  {A.}~\bibnamefont {Alù}},\ and\ \bibinfo {author} {\bibfnamefont {A.~B.}\
  \bibnamefont {Khanikaev}},\ }\bibfield  {title} {\bibinfo {title} {{Optical
  isolator based on chiral light-matter interactions in a ring resonator
  integrating a dichroic magneto-optical material}},\ }\href
  {https://doi.org/10.1063/5.0057558} {\bibfield  {journal} {\bibinfo
  {journal} {Applied Physics Letters}\ }\textbf {\bibinfo {volume} {118}},\
  \bibinfo {pages} {241104} (\bibinfo {year} {2021})}\BibitemShut {NoStop}%
\bibitem [{\citenamefont {Boddeti}\ \emph {et~al.}(2024)\citenamefont
  {Boddeti}, \citenamefont {Wang}, \citenamefont {Juarez}, \citenamefont
  {Boltasseva}, \citenamefont {Odom}, \citenamefont {Shalaev}, \citenamefont
  {Alaeian},\ and\ \citenamefont {Jacob}}]{Reduced_dimension_Boddeti2024}%
  \BibitemOpen
  \bibfield  {author} {\bibinfo {author} {\bibfnamefont {A.~K.}\ \bibnamefont
  {Boddeti}}, \bibinfo {author} {\bibfnamefont {Y.}~\bibnamefont {Wang}},
  \bibinfo {author} {\bibfnamefont {X.~G.}\ \bibnamefont {Juarez}}, \bibinfo
  {author} {\bibfnamefont {A.}~\bibnamefont {Boltasseva}}, \bibinfo {author}
  {\bibfnamefont {T.~W.}\ \bibnamefont {Odom}}, \bibinfo {author}
  {\bibfnamefont {V.}~\bibnamefont {Shalaev}}, \bibinfo {author} {\bibfnamefont
  {H.}~\bibnamefont {Alaeian}},\ and\ \bibinfo {author} {\bibfnamefont
  {Z.}~\bibnamefont {Jacob}},\ }\bibfield  {title} {\bibinfo {title} {Reducing
  effective system dimensionality with long-range collective dipole-dipole
  interactions},\ }\href {https://doi.org/10.1103/PhysRevLett.132.173803}
  {\bibfield  {journal} {\bibinfo  {journal} {Physical Review Letters}\
  }\textbf {\bibinfo {volume} {132}},\ \bibinfo {pages} {173803} (\bibinfo
  {year} {2024})}\BibitemShut {NoStop}%
\bibitem [{\citenamefont {Ashida}\ \emph {et~al.}(2020)\citenamefont {Ashida},
  \citenamefont {Imamoglu}, \citenamefont {Faist}, \citenamefont {Jaksch},
  \citenamefont {Cavalleri},\ and\ \citenamefont {Demler}}]{Ashida2020}%
  \BibitemOpen
  \bibfield  {author} {\bibinfo {author} {\bibfnamefont {Y.}~\bibnamefont
  {Ashida}}, \bibinfo {author} {\bibfnamefont {A.}~\bibnamefont {Imamoglu}},
  \bibinfo {author} {\bibfnamefont {J.}~\bibnamefont {Faist}}, \bibinfo
  {author} {\bibfnamefont {D.}~\bibnamefont {Jaksch}}, \bibinfo {author}
  {\bibfnamefont {A.}~\bibnamefont {Cavalleri}},\ and\ \bibinfo {author}
  {\bibfnamefont {E.}~\bibnamefont {Demler}},\ }\bibfield  {title} {\bibinfo
  {title} {Quantum electrodynamics in two-dimensional materials},\ }\href@noop
  {} {\bibfield  {journal} {\bibinfo  {journal} {Phys. Rev. X}\ }\textbf
  {\bibinfo {volume} {10}},\ \bibinfo {pages} {041027} (\bibinfo {year}
  {2020})}\BibitemShut {NoStop}%
\bibitem [{\citenamefont {Sentef}\ \emph {et~al.}(2018)\citenamefont {Sentef},
  \citenamefont {Ruggenthaler},\ and\ \citenamefont {Rubio}}]{Sentef2018}%
  \BibitemOpen
  \bibfield  {author} {\bibinfo {author} {\bibfnamefont {M.~A.}\ \bibnamefont
  {Sentef}}, \bibinfo {author} {\bibfnamefont {M.}~\bibnamefont
  {Ruggenthaler}},\ and\ \bibinfo {author} {\bibfnamefont {A.}~\bibnamefont
  {Rubio}},\ }\bibfield  {title} {\bibinfo {title} {Cavity
  quantum-electrodynamical purcell effect in 2d materials},\ }\href@noop {}
  {\bibfield  {journal} {\bibinfo  {journal} {Science Advances}\ }\textbf
  {\bibinfo {volume} {4}},\ \bibinfo {pages} {eaau6969} (\bibinfo {year}
  {2018})}\BibitemShut {NoStop}%
\bibitem [{\citenamefont {Curtis}\ \emph {et~al.}(2019)\citenamefont {Curtis},
  \citenamefont {Raines}, \citenamefont {Allocca}, \citenamefont {Hafezi},\
  and\ \citenamefont {Galitski}}]{Curtis2019}%
  \BibitemOpen
  \bibfield  {author} {\bibinfo {author} {\bibfnamefont {J.~B.}\ \bibnamefont
  {Curtis}}, \bibinfo {author} {\bibfnamefont {Z.~M.}\ \bibnamefont {Raines}},
  \bibinfo {author} {\bibfnamefont {A.~A.}\ \bibnamefont {Allocca}}, \bibinfo
  {author} {\bibfnamefont {M.}~\bibnamefont {Hafezi}},\ and\ \bibinfo {author}
  {\bibfnamefont {V.~M.}\ \bibnamefont {Galitski}},\ }\bibfield  {title}
  {\bibinfo {title} {Cavity quantum electrodynamics with two-dimensional
  materials},\ }\href@noop {} {\bibfield  {journal} {\bibinfo  {journal} {Phys.
  Rev. Lett.}\ }\textbf {\bibinfo {volume} {122}},\ \bibinfo {pages} {167002}
  (\bibinfo {year} {2019})}\BibitemShut {NoStop}%
\bibitem [{\citenamefont {Schlawin}\ \emph {et~al.}(2019)\citenamefont
  {Schlawin}, \citenamefont {Cavalleri},\ and\ \citenamefont
  {Jaksch}}]{Schlawin2019}%
  \BibitemOpen
  \bibfield  {author} {\bibinfo {author} {\bibfnamefont {F.}~\bibnamefont
  {Schlawin}}, \bibinfo {author} {\bibfnamefont {A.}~\bibnamefont
  {Cavalleri}},\ and\ \bibinfo {author} {\bibfnamefont {D.}~\bibnamefont
  {Jaksch}},\ }\bibfield  {title} {\bibinfo {title} {Cavity-mediated
  electron-photon superconductivity},\ }\href@noop {} {\bibfield  {journal}
  {\bibinfo  {journal} {Phys. Rev. Lett.}\ }\textbf {\bibinfo {volume} {122}},\
  \bibinfo {pages} {133602} (\bibinfo {year} {2019})}\BibitemShut {NoStop}%
\bibitem [{\citenamefont {Latini}\ \emph {et~al.}(2021)\citenamefont {Latini},
  \citenamefont {Shin}, \citenamefont {Sato}, \citenamefont {Sch\"afer},
  \citenamefont {Giovannini}, \citenamefont {H\"ubener},\ and\ \citenamefont
  {Rubio}}]{Latini2021}%
  \BibitemOpen
  \bibfield  {author} {\bibinfo {author} {\bibfnamefont {S.}~\bibnamefont
  {Latini}}, \bibinfo {author} {\bibfnamefont {D.}~\bibnamefont {Shin}},
  \bibinfo {author} {\bibfnamefont {S.~A.}\ \bibnamefont {Sato}}, \bibinfo
  {author} {\bibfnamefont {C.}~\bibnamefont {Sch\"afer}}, \bibinfo {author}
  {\bibfnamefont {U.~D.}\ \bibnamefont {Giovannini}}, \bibinfo {author}
  {\bibfnamefont {H.}~\bibnamefont {H\"ubener}},\ and\ \bibinfo {author}
  {\bibfnamefont {A.}~\bibnamefont {Rubio}},\ }\bibfield  {title} {\bibinfo
  {title} {Cavity control of excitons in two-dimensional materials},\
  }\href@noop {} {\bibfield  {journal} {\bibinfo  {journal} {Proceedings of the
  National Academy of Sciences}\ }\textbf {\bibinfo {volume} {118}},\ \bibinfo
  {pages} {e2105618118} (\bibinfo {year} {2021})}\BibitemShut {NoStop}%
\bibitem [{\citenamefont {Mivehvar}\ \emph {et~al.}(2017)\citenamefont
  {Mivehvar}, \citenamefont {Ritsch},\ and\ \citenamefont
  {Piazza}}]{Mivehvar2017}%
  \BibitemOpen
  \bibfield  {author} {\bibinfo {author} {\bibfnamefont {F.}~\bibnamefont
  {Mivehvar}}, \bibinfo {author} {\bibfnamefont {H.}~\bibnamefont {Ritsch}},\
  and\ \bibinfo {author} {\bibfnamefont {F.}~\bibnamefont {Piazza}},\
  }\bibfield  {title} {\bibinfo {title} {Superradiant topological peierls
  insulator},\ }\href@noop {} {\bibfield  {journal} {\bibinfo  {journal} {Phys.
  Rev. Lett.}\ }\textbf {\bibinfo {volume} {118}},\ \bibinfo {pages} {073602}
  (\bibinfo {year} {2017})}\BibitemShut {NoStop}%
\bibitem [{\citenamefont {Dmytruk}\ and\ \citenamefont
  {Schir{\`o}}(2022)}]{Dmytruk2022}%
  \BibitemOpen
  \bibfield  {author} {\bibinfo {author} {\bibfnamefont {O.}~\bibnamefont
  {Dmytruk}}\ and\ \bibinfo {author} {\bibfnamefont {M.}~\bibnamefont
  {Schir{\`o}}},\ }\bibfield  {title} {\bibinfo {title} {Controlling
  topological phases of matter with quantum light},\ }\href
  {https://doi.org/10.1038/s42005-022-01070-7} {\bibfield  {journal} {\bibinfo
  {journal} {Communications Physics}\ }\textbf {\bibinfo {volume} {5}},\
  \bibinfo {pages} {Article 271} (\bibinfo {year} {2022})}\BibitemShut
  {NoStop}%
\bibitem [{\citenamefont {Appugliese}\ \emph {et~al.}(2022)\citenamefont
  {Appugliese}, \citenamefont {Enkner}, \citenamefont {Paravicini-Bagliani},
  \citenamefont {Beck}, \citenamefont {Reichl}, \citenamefont {Wegscheider},
  \citenamefont {Scalari}, \citenamefont {Ciuti},\ and\ \citenamefont
  {Faist}}]{Appugliese2022}%
  \BibitemOpen
  \bibfield  {author} {\bibinfo {author} {\bibfnamefont {F.}~\bibnamefont
  {Appugliese}}, \bibinfo {author} {\bibfnamefont {J.}~\bibnamefont {Enkner}},
  \bibinfo {author} {\bibfnamefont {G.~L.}\ \bibnamefont
  {Paravicini-Bagliani}}, \bibinfo {author} {\bibfnamefont {M.}~\bibnamefont
  {Beck}}, \bibinfo {author} {\bibfnamefont {C.}~\bibnamefont {Reichl}},
  \bibinfo {author} {\bibfnamefont {W.}~\bibnamefont {Wegscheider}}, \bibinfo
  {author} {\bibfnamefont {G.}~\bibnamefont {Scalari}}, \bibinfo {author}
  {\bibfnamefont {C.}~\bibnamefont {Ciuti}},\ and\ \bibinfo {author}
  {\bibfnamefont {J.}~\bibnamefont {Faist}},\ }\bibfield  {title} {\bibinfo
  {title} {Breakdown of topological protection by cavity vacuum fields in the
  integer quantum hall effect},\ }\href@noop {} {\bibfield  {journal} {\bibinfo
   {journal} {Science}\ }\textbf {\bibinfo {volume} {375}},\ \bibinfo {pages}
  {1030} (\bibinfo {year} {2022})}\BibitemShut {NoStop}%
\bibitem [{\citenamefont {Bacciconi}\ \emph {et~al.}(2023)\citenamefont
  {Bacciconi}, \citenamefont {Andolina}, \citenamefont {Chanda}, \citenamefont
  {Chiriacò}, \citenamefont {Schirò},\ and\ \citenamefont
  {Dalmonte}}]{Bacciconi2023}%
  \BibitemOpen
  \bibfield  {author} {\bibinfo {author} {\bibfnamefont {Z.}~\bibnamefont
  {Bacciconi}}, \bibinfo {author} {\bibfnamefont {G.~M.}\ \bibnamefont
  {Andolina}}, \bibinfo {author} {\bibfnamefont {T.}~\bibnamefont {Chanda}},
  \bibinfo {author} {\bibfnamefont {G.}~\bibnamefont {Chiriacò}}, \bibinfo
  {author} {\bibfnamefont {M.}~\bibnamefont {Schirò}},\ and\ \bibinfo {author}
  {\bibfnamefont {M.}~\bibnamefont {Dalmonte}},\ }\bibfield  {title} {\bibinfo
  {title} {Photon-mediated electron pairing in cavity quantum
  electrodynamics},\ }\href@noop {} {\bibfield  {journal} {\bibinfo  {journal}
  {SciPost Phys.}\ }\textbf {\bibinfo {volume} {15}},\ \bibinfo {pages} {113}
  (\bibinfo {year} {2023})}\BibitemShut {NoStop}%
\bibitem [{\citenamefont {Chiocchetta}\ \emph {et~al.}(2021)\citenamefont
  {Chiocchetta}, \citenamefont {Kiese}, \citenamefont {Zelle}, \citenamefont
  {Piazza},\ and\ \citenamefont {Diehl}}]{Chiocchetta2021}%
  \BibitemOpen
  \bibfield  {author} {\bibinfo {author} {\bibfnamefont {A.}~\bibnamefont
  {Chiocchetta}}, \bibinfo {author} {\bibfnamefont {D.}~\bibnamefont {Kiese}},
  \bibinfo {author} {\bibfnamefont {C.~P.}\ \bibnamefont {Zelle}}, \bibinfo
  {author} {\bibfnamefont {F.}~\bibnamefont {Piazza}},\ and\ \bibinfo {author}
  {\bibfnamefont {S.}~\bibnamefont {Diehl}},\ }\bibfield  {title} {\bibinfo
  {title} {Cavity-induced quantum spin liquids},\ }\href@noop {} {\bibfield
  {journal} {\bibinfo  {journal} {Nature Communications}\ }\textbf {\bibinfo
  {volume} {12}},\ \bibinfo {pages} {5901} (\bibinfo {year}
  {2021})}\BibitemShut {NoStop}%
\bibitem [{\citenamefont {Mercurio}\ \emph {et~al.}(2024)\citenamefont
  {Mercurio}, \citenamefont {Andolina}, \citenamefont {Pellegrino},
  \citenamefont {Di~Stefano}, \citenamefont {Jarillo-Herrero}, \citenamefont
  {Felser}, \citenamefont {Koppens}, \citenamefont {Savasta},\ and\
  \citenamefont {Polini}}]{Mercurio2023}%
  \BibitemOpen
  \bibfield  {author} {\bibinfo {author} {\bibfnamefont {A.}~\bibnamefont
  {Mercurio}}, \bibinfo {author} {\bibfnamefont {G.~M.}\ \bibnamefont
  {Andolina}}, \bibinfo {author} {\bibfnamefont {F.~M.~D.}\ \bibnamefont
  {Pellegrino}}, \bibinfo {author} {\bibfnamefont {O.}~\bibnamefont
  {Di~Stefano}}, \bibinfo {author} {\bibfnamefont {P.}~\bibnamefont
  {Jarillo-Herrero}}, \bibinfo {author} {\bibfnamefont {C.}~\bibnamefont
  {Felser}}, \bibinfo {author} {\bibfnamefont {F.~H.~L.}\ \bibnamefont
  {Koppens}}, \bibinfo {author} {\bibfnamefont {S.}~\bibnamefont {Savasta}},\
  and\ \bibinfo {author} {\bibfnamefont {M.}~\bibnamefont {Polini}},\
  }\bibfield  {title} {\bibinfo {title} {Photon condensation, van vleck
  paramagnetism, and chiral cavities},\ }\href
  {https://doi.org/10.1103/PhysRevResearch.6.013303} {\bibfield  {journal}
  {\bibinfo  {journal} {Phys. Rev. Res.}\ }\textbf {\bibinfo {volume} {6}},\
  \bibinfo {pages} {013303} (\bibinfo {year} {2024})}\BibitemShut {NoStop}%
\bibitem [{Sup()}]{SupplementaryMaterial}%
  \BibitemOpen
  \href@noop {} {}\bibinfo {note} {Supplemental Material}\BibitemShut {NoStop}%
\bibitem [{\citenamefont {Sadzi}\ and\ \citenamefont
  {Hadad}(2024)}]{koffi2024}%
  \BibitemOpen
  \bibfield  {author} {\bibinfo {author} {\bibfnamefont {K.-E.}\ \bibnamefont
  {Sadzi}}\ and\ \bibinfo {author} {\bibfnamefont {Y.}~\bibnamefont {Hadad}},\
  }\bibfield  {title} {\bibinfo {title} {The mutual dynamics of a resonant
  particle inside rectangular cavity: Collective polarizability calculation via
  a ladder-type alternative green's functions approach}} (\bibinfo {year}
  {2024}),\ \bibinfo {note} {preprint}\BibitemShut {NoStop}%
\bibitem [{\citenamefont {Novotny}\ and\ \citenamefont
  {Hecht}(2012)}]{novotny2012principles}%
  \BibitemOpen
  \bibfield  {author} {\bibinfo {author} {\bibfnamefont {L.}~\bibnamefont
  {Novotny}}\ and\ \bibinfo {author} {\bibfnamefont {B.}~\bibnamefont
  {Hecht}},\ }\href {https://www.cambridge.org/9781107005464} {\emph {\bibinfo
  {title} {Principles of Nano-Optics}}},\ \bibinfo {edition} {second edition}\
  ed.\ (\bibinfo  {publisher} {Cambridge University Press},\ \bibinfo {address}
  {Cambridge, New York, Melbourne, Madrid, Cape Town, Singapore, S{\~a}o Paulo,
  Delhi, Mexico City},\ \bibinfo {year} {2012})\BibitemShut {NoStop}%
\bibitem [{\citenamefont {Collin}(1991)}]{collin1991field}%
  \BibitemOpen
  \bibfield  {author} {\bibinfo {author} {\bibfnamefont {R.~E.}\ \bibnamefont
  {Collin}},\ }\href@noop {} {\emph {\bibinfo {title} {Field Theory of Guided
  Waves}}},\ \bibinfo {edition} {2nd}\ ed.\ (\bibinfo  {publisher}
  {McGraw-Hill, IEEE},\ \bibinfo {year} {1991})\BibitemShut {NoStop}%
\bibitem [{\citenamefont {Rumsey}(1954)}]{rumsey1954reaction}%
  \BibitemOpen
  \bibfield  {author} {\bibinfo {author} {\bibfnamefont {V.~H.}\ \bibnamefont
  {Rumsey}},\ }\bibfield  {title} {\bibinfo {title} {Reaction concept in
  electromagnetic theory},\ }\href@noop {} {\bibfield  {journal} {\bibinfo
  {journal} {Physical Review}\ }\textbf {\bibinfo {volume} {94}},\ \bibinfo
  {pages} {1483} (\bibinfo {year} {1954})}\BibitemShut {NoStop}%
\bibitem [{\citenamefont {Kong}(1972)}]{kong1972theorems}%
  \BibitemOpen
  \bibfield  {author} {\bibinfo {author} {\bibfnamefont {J.~A.}\ \bibnamefont
  {Kong}},\ }\bibfield  {title} {\bibinfo {title} {Theorems of bianisotropic
  media},\ }\href@noop {} {\bibfield  {journal} {\bibinfo  {journal}
  {Proceedings of the IEEE}\ }\textbf {\bibinfo {volume} {60}},\ \bibinfo
  {pages} {1036} (\bibinfo {year} {1972})}\BibitemShut {NoStop}%
\bibitem [{\citenamefont {Rodríguez}\ \emph {et~al.}(2013)\citenamefont
  {Rodríguez}, \citenamefont {Martínez}, \citenamefont {Rojas},\ and\
  \citenamefont {Querts}}]{Rodriguez2013}%
  \BibitemOpen
  \bibfield  {author} {\bibinfo {author} {\bibfnamefont {L.~C.}\ \bibnamefont
  {Rodríguez}}, \bibinfo {author} {\bibfnamefont {A.~P.}\ \bibnamefont
  {Martínez}}, \bibinfo {author} {\bibfnamefont {H.~P.}\ \bibnamefont
  {Rojas}},\ and\ \bibinfo {author} {\bibfnamefont {E.~R.}\ \bibnamefont
  {Querts}},\ }\bibfield  {title} {\bibinfo {title} {Quantized faraday effect
  in (3 + 1)-dimensional and (2 + 1)-dimensional systems},\ }\href
  {https://doi.org/10.1103/PhysRevA.88.052126} {\bibfield  {journal} {\bibinfo
  {journal} {Physical Review A}\ }\textbf {\bibinfo {volume} {88}},\ \bibinfo
  {pages} {052126} (\bibinfo {year} {2013})}\BibitemShut {NoStop}%
\bibitem [{\citenamefont {Burla}\ \emph {et~al.}(2023)\citenamefont {Burla},
  \citenamefont {Hoessbacher}, \citenamefont {Heni}, \citenamefont {Haffner},
  \citenamefont {Salamin}, \citenamefont {Fedoryshyn}, \citenamefont
  {Watanabe}, \citenamefont {Massler}, \citenamefont {Blatter}, \citenamefont
  {Horst}, \citenamefont {Elder}, \citenamefont {Dalton},\ and\ \citenamefont
  {Leuthold}}]{burla2023}%
  \BibitemOpen
  \bibfield  {author} {\bibinfo {author} {\bibfnamefont {M.}~\bibnamefont
  {Burla}}, \bibinfo {author} {\bibfnamefont {C.}~\bibnamefont {Hoessbacher}},
  \bibinfo {author} {\bibfnamefont {W.}~\bibnamefont {Heni}}, \bibinfo {author}
  {\bibfnamefont {C.}~\bibnamefont {Haffner}}, \bibinfo {author} {\bibfnamefont
  {Y.}~\bibnamefont {Salamin}}, \bibinfo {author} {\bibfnamefont
  {Y.}~\bibnamefont {Fedoryshyn}}, \bibinfo {author} {\bibfnamefont
  {T.}~\bibnamefont {Watanabe}}, \bibinfo {author} {\bibfnamefont
  {H.}~\bibnamefont {Massler}}, \bibinfo {author} {\bibfnamefont
  {T.}~\bibnamefont {Blatter}}, \bibinfo {author} {\bibfnamefont
  {Y.}~\bibnamefont {Horst}}, \bibinfo {author} {\bibfnamefont {D.~L.}\
  \bibnamefont {Elder}}, \bibinfo {author} {\bibfnamefont {L.~R.}\ \bibnamefont
  {Dalton}},\ and\ \bibinfo {author} {\bibfnamefont {J.}~\bibnamefont
  {Leuthold}},\ }\bibfield  {title} {\bibinfo {title} {Plasmonics for microwave
  photonics in the thz range},\ }\href
  {https://doi.org/10.3389/fphot.2023.1067916} {\bibfield  {journal} {\bibinfo
  {journal} {Frontiers in Photonics}\ }\textbf {\bibinfo {volume} {4}},\
  \bibinfo {pages} {1067916} (\bibinfo {year} {2023})}\BibitemShut {NoStop}%
\bibitem [{\citenamefont {Haffner}\ \emph {et~al.}(2017)\citenamefont
  {Haffner}, \citenamefont {Heni}, \citenamefont {Elder}, \citenamefont
  {Fedoryshyn}, \citenamefont {Đorđević}, \citenamefont {Chelladurai},
  \citenamefont {Koch}, \citenamefont {Portner}, \citenamefont {Burla},
  \citenamefont {Robinson}, \citenamefont {Dalton},\ and\ \citenamefont
  {Leuthold}}]{Haffner2017}%
  \BibitemOpen
  \bibfield  {author} {\bibinfo {author} {\bibfnamefont {C.}~\bibnamefont
  {Haffner}}, \bibinfo {author} {\bibfnamefont {W.}~\bibnamefont {Heni}},
  \bibinfo {author} {\bibfnamefont {D.~L.}\ \bibnamefont {Elder}}, \bibinfo
  {author} {\bibfnamefont {Y.}~\bibnamefont {Fedoryshyn}}, \bibinfo {author}
  {\bibfnamefont {N.}~\bibnamefont {Đorđević}}, \bibinfo {author}
  {\bibfnamefont {D.}~\bibnamefont {Chelladurai}}, \bibinfo {author}
  {\bibfnamefont {U.}~\bibnamefont {Koch}}, \bibinfo {author} {\bibfnamefont
  {K.}~\bibnamefont {Portner}}, \bibinfo {author} {\bibfnamefont
  {M.}~\bibnamefont {Burla}}, \bibinfo {author} {\bibfnamefont
  {B.}~\bibnamefont {Robinson}}, \bibinfo {author} {\bibfnamefont {L.~R.}\
  \bibnamefont {Dalton}},\ and\ \bibinfo {author} {\bibfnamefont
  {J.}~\bibnamefont {Leuthold}},\ }\bibfield  {title} {\bibinfo {title}
  {Harnessing nonlinearities near material absorption resonances for reducing losses in plasmonic modulators},\ }\href
  {https://doi.org/10.1364/OME.7.002168} {\bibfield  {journal} {\bibinfo
  {journal} {Optical Materials Express}\ }\textbf {\bibinfo {volume} {7}},\
  \bibinfo {pages} {2168} (\bibinfo {year} {2017})}\BibitemShut {NoStop}%
\bibitem [{\citenamefont {Scheuer}\ and\ \citenamefont
  {Yariv}(2006)}]{Yariv_sagnac2006}%
  \BibitemOpen
  \bibfield  {author} {\bibinfo {author} {\bibfnamefont {J.}~\bibnamefont
  {Scheuer}}\ and\ \bibinfo {author} {\bibfnamefont {A.}~\bibnamefont
  {Yariv}},\ }\bibfield  {title} {\bibinfo {title} {Sagnac effect in
  coupled-resonator slow-light waveguide structures},\ }\href
  {https://doi.org/10.1103/PhysRevLett.96.053901} {\bibfield  {journal}
  {\bibinfo  {journal} {Phys. Rev. Lett.}\ }\textbf {\bibinfo {volume} {96}},\
  \bibinfo {pages} {053901} (\bibinfo {year} {2006})}\BibitemShut {NoStop}%
\bibitem [{\citenamefont {Novitski}\ \emph {et~al.}(2012)\citenamefont
  {Novitski}, \citenamefont {Steinberg},\ and\ \citenamefont
  {Scheuer}}]{Steinberg_CROW2012}%
  \BibitemOpen
  \bibfield  {author} {\bibinfo {author} {\bibfnamefont {R.}~\bibnamefont
  {Novitski}}, \bibinfo {author} {\bibfnamefont {B.~Z.}\ \bibnamefont
  {Steinberg}},\ and\ \bibinfo {author} {\bibfnamefont {J.}~\bibnamefont
  {Scheuer}},\ }\bibfield  {title} {\bibinfo {title} {Losses in rotating
  degenerate cavities and a coupled-resonator optical-waveguide rotation
  sensor},\ }\href {https://doi.org/10.1103/PhysRevA.85.023813} {\bibfield
  {journal} {\bibinfo  {journal} {Phys. Rev. A}\ }\textbf {\bibinfo {volume}
  {85}},\ \bibinfo {pages} {023813} (\bibinfo {year} {2012})}\BibitemShut
  {NoStop}%
\end{thebibliography}

\begin{thebibliography}{10}
\bibitem{ishimaru1991electromagnetic}
A.~Ishimaru, \textit{Electromagnetic Wave Propagation, Radiation, and Scattering} (Prentice Hall, Englewood Cliffs, NJ, 1991). ISBN: 978-0132490535.

\bibitem{koffi2024}
K.-E.~Sadzi and Y.~Hadad, ``The mutual dynamics of a resonant particle inside rectangular cavity: Collective polarizability calculation via a ladder-type alternative Green's functions approach,'' preprint (2024).

\bibitem{novotny2012principles}
L.~Novotny and B.~Hecht, \textit{Principles of Nano-Optics}, 2nd ed. (Cambridge University Press, Cambridge, 2012). ISBN: 978-1-107-00546-4.

\bibitem{rumsey1954reaction}
V.~H. Rumsey, ``Reaction Concept in Electromagnetic Theory,'' \textit{Physical Review} \textbf{94}, 1483--1491 (1954).

\bibitem{kong1972theorems}
J.~A. Kong, ``Theorems of Bianisotropic Media,'' \textit{Proceedings of the IEEE} \textbf{60}, 1036--1046 (1972).

\bibitem{collin1991field}
R.~E. Collin, \textit{Field Theory of Guided Waves}, 2nd ed. (McGraw-Hill, IEEE, 1991). ISBN: 978-0-070-10044-9.
\end{thebibliography}

\providecommand{\noopsort}[1]{}\providecommand{\singleletter}[1]{#1}%

\newpage
\onecolumngrid

SUPPLEMENTAL MATERIAL

\renewcommand{\theequation}{S\arabic{equation}}

\section{Dynamic Polarizability tensor of a gyrotropic sphere \label{polarizability_tensor}}

The dynamic polarizability tensor of a spherical magnetized plasmonic particle  that is characterized by a gyrotropic permittivity model  \cite{ishimaru1991electromagnetic} can be expressed as,  
\begin{subequations}
\label{dynamic_polz}
\begin{align}
    & \underline{\underline{\alpha}}_e^{-1} = \frac{k^3}{6 \pi \varepsilon_0 } 
   \left( 
 \underline{\underline{\alpha}_h}^{-1}
    +j \underline{\underline{I}}\right) \\
    &\underline{\underline{\alpha}_h} ^{-1}=
    \begin{bmatrix}
        g_{xx} & -j g_{xy} & 0\\
        j g_{xy} & g_{yy} & 0 \\
        0 &  0 & g_{zz}
    \end{bmatrix}
\end{align}
with 

\begin{align}
&g_{uu} = \frac{6\pi}{k^3 V }  \left( \frac{1}{3} - \frac{\omega (\omega - 2\pi j\Gamma )}{\omega_p^2} \cdot \frac{(\omega - 2\pi j\Gamma)^2 -\omega_c^2}{(\omega - 2\pi j\Gamma)^2 +\omega_c^2} \right)
\\
&g_{zz} = \frac{6\pi}{k^3 V }  \left( \frac{1}{3} - \frac{\omega (\omega - 2\pi j\Gamma )}{\omega_p^2} \right)
\\
&g_{xy} =  \frac{6\pi}{k^3 V } \left[ \frac{\omega_c (\omega - 2\pi j\Gamma)}{\omega_p^2} \cdot \frac{(\omega - 2\pi j\Gamma)^2 -\omega_c^2}{(\omega - 2\pi j\Gamma)^2 
+\omega_c^2}\right.  
-\left. \frac{2}{3} \frac{\omega_c (\omega - 2\pi j\Gamma)}{(\omega - 2\pi j\Gamma)^2 +\omega_c^2} \right]
\end{align}
\end{subequations}
where $\omega_p$, $\omega_c=-q_e B_0/m_e$ and $2\pi \Gamma$ are the plasma, cyclotron, and collision frequencies, respectively.
In the absence of loss, $\Gamma=0$, and $\underline{\underline{\alpha}}_h^{-1}$ is hermitian. 

Inside a complex domain (i.e., excluding the free space), the effective (also termed collective) dynamic polarizability tensor is given  by \cite{koffi2024,novotny2012principles}
\begin{align}   
\label{alfa_eff}
\underline{\underline{\alpha}_{\rm eff}}^{-1}= \left[\underline{\underline{\alpha}}_e ^{-1}-{\underline{\underline{G}}^{\rm loc}(\bm{r}',\bm{r}',\omega)} \right]
\end{align}
where ${\underline{\underline{G}}}^{\rm loc}$ denotes the dyadic Green function of the local electric field at the location of the particle $\bm{r'}$. We derive this dyadic Green's function for a resonant particle inside the rectangular cavity in \cite{koffi2024}.
In free space, ${\underline{\underline{G}}}^{\rm loc}=0$ and one gets back
$\underline{\underline{\alpha}}_{\rm eff}=\underline{\underline{\alpha}_{e}}$. In general, in order to find the system resonances the determinant of the inverse of the collective polarizability is set to zero, i.e., $\mbox{det}\left\{\underline{\underline{\alpha}}_{\rm eff}^{-1}\right\}=0$. 
To find the excitation of the gyrotropic sphere due to one of the testing sources, say $\bm{p}_1$, we solved,
\begin{equation}
\label{pg_eq}
\bm{p}_g=\underline{\underline{\alpha}}_e \underline{\underline{G}}^{\rm loc}(\bm{r}_p,\bm{r}_p)\bm{p}_{g} + \underline{\underline{\alpha}}_e \underline{\underline{G}}_{e}(\bm{r}_p,\bm{r}_1) \bm{p}_1
\end{equation}
where $\underline{\underline{G}}_{e}$ is the cavity Green's function that is given in App.~\ref{cavity_GF}.
Eq.~(\ref{pg_eq}) yields:
\begin{equation}
    \bm{p}_g=\underline{\underline{\alpha}}_{\rm eff} \underline{\underline{G}}_{e}(\bm{r}_p,\bm{r}_1) \bm{p}_1
\end{equation}
where $\underline{\underline{\alpha}}_{\rm eff}$ is given in Eq.~(\ref{alfa_eff}).

Once $\bm{p}_g$ is found, the field $\bm{E}_1$ at $\bm{r}_2$ due to $\bm{p}_1$ reads, 
\begin{equation}
\bm{E}_1=\underline{\underline{G}}_{e}(\bm{r}_2,\bm{r}_g)\bm{p}_g +\underline{\underline{G}}_{e}(\bm{r}_2,\bm{r}_1)\bm{p}_1.  
\end{equation}

% It is known that the imaginary part of $\underline{\underline{\alpha_e}} ^{-1}$ and $\bm{\underline{\underline{G}}}^{loc}$ is equal to the electric radiation correction term \cite{}, and the latter is a symmetric matrix, and we get that $ \underline{\underline{\alpha_{eff}}}^{-1}$ is also hermitian for $\Gamma=0$.

\section{ \label{Lorent_coef} Lorentz reciprocity coefficient and its properties inside the rectangular cavity box }

Lorentz reciprocity theorem  can be expressed as follows:
\begin{equation}
\label{lorentz1}
\int_V (\bm{J}_1 \cdot \bm{E}_2 - \bm{J}_2 \cdot \bm{E}_1) \, dV = 0
\end{equation}
where $\bm{E}_1$ and $\bm{E}_2$ are the electric fields produced by the current densities $\bm{J}_1$ and $\bm{J}_2$, respectively, and $V$ is the volume over which the integration is performed. 

Using the definition of reaction \cite{rumsey1954reaction,kong1972theorems}
\begin{equation}
    <1,2>= \int_V \bm{J}_1 \cdot \bm{E}_2 \, dV 
\end{equation}
Eq.~(\ref{lorentz1}) can be rewritten as
\begin{equation}
    <1,2>-<2,1>=0
\end{equation}
For $\bm{J}_i=j\omega \, \bm{p}_i  \delta(\bm{r}-\bm{r}')$, Eq.~(\ref{lorentz1}) becomes
\begin{equation}
\label{lorentz2}
 ( \bm{p}_1 \cdot \bm{E}_2 -  \bm{p}_2 \cdot \bm{E}_1) = 0
\end{equation}
where the factor $j\omega$ is omitted.

We hence define the reciprocity measure (Eq.~(1) in the main text)
\begin{align}
    \mathcal{R}=\frac{\bm{p}_1 \cdot \bm{E}_2- \bm{p}_2 \cdot \bm{E}_1 }{|\bm{p}_1 \cdot \bm{E}_2|+ |\bm{p}_2 \cdot \bm{E}_1|}
\end{align}

We define the medium $\mathcal{M}$ inside the rectangular cavity containing a spherical gyrotropic particle with effective polarizability $ \underline{\underline{\alpha}}_{\rm eff}$ located at $\bm{r}_g$. We can use the cavity dyadic Green function to express $\mathcal{R}$ for two radiating testing electric dipoles $\bm{p}_1$ and $\bm{p}_2$ located at $\bm{r}_1$ and $\bm{r}_2$, respectively, inside $\mathcal{M}$.
The reactions are expressed as follows:

\begin{align}
\label{firstpart_lorentz}
    {\bm{p}_2}\cdot \bm{E}_1=&  
    \bm{p}_2^T \underline{\underline{G}}_e(\bm{r}_2,\bm{r}_1)\bm{p}_1  +  
    \bm{p}_2^T\left[ \underline{\underline{G}}_e(\bm{r}_2,\bm{r}_g)  \,\underline{\underline{\alpha}}_{\rm eff}   \,\underline{\underline{G}}_e(\bm{r}_g,\bm{r}_1)
     \right]\bm{p}_1
\end{align}

\begin{align}
\label{2ndpart_lorentz}
    {\bm{p}_1}\cdot \bm{E}_2 =& \bm{p}_1^T  \underline{\underline{G}}_e(\bm{r}_1,\bm{r}_2) \bm{p}_2 + \bm{p}_1^T \left[
     \underline{\underline{G}}_e(\bm{r}_1,\bm{r}_g)  \,\underline{\underline{\alpha}}_{\rm eff}   \,\underline{\underline{G}}_e(\bm{r}_g,\bm{r}_2) 
     \right] \bm{p}_2 
\end{align}
Using the reciprocity of the Green function inside the empty cavity, and the reaction being a scalar one easily gets,
\begin{align}
     {\bm{p}_1}\cdot \bm{E}_2 =& \bm{p}_2^T \underline{\underline{G}}_e(\bm{r}_2,\bm{r}_1) \bm{p}_1 + \bm{p}_2^T  \left[ \,   \underline{\underline{G}}_e(\bm{r}_2,\bm{r}_g)\,\underline{\underline{\alpha}}^T_{\rm eff}   \,\underline{\underline{G}}_e(\bm{r}_g,\bm{r}_2) \right]  \, \bm{p}_1
\end{align}
For real-valued $\bm{p}_1$ and $\bm{p}_2$  (this does not limit the generality as an additional phase just implies a relocation of the testing dipoles inside the cavity) and using the fact that the Green's function $\underline{\underline{G}}_e(\bm{r},\bm{r}')$ is real-valued (see App.~\ref{cavity_GF})
\begin{align}
\label{conj_1}
    \mbox{Im}\{ {\bm{p}_1}\cdot \bm{E}_2\} &=\bm{p}_2^T   \,   \underline{\underline{G}}_e(\bm{r}_2,\bm{r}_g)\,\mbox{Im}\{\underline{\underline{\alpha}}^T_{\rm eff}  \} \,\underline{\underline{G}}_e(\bm{r}_g,\bm{r}_2)   \, \bm{p}_1  \nonumber
    \\&=
    -\bm{p}_2^T   \,   \underline{\underline{G}}_e(\bm{r}_2,\bm{r}_g)\,\mbox{Im}\{\underline{\underline{\alpha}}_{\rm eff}  \} \,\underline{\underline{G}}_e(\bm{r}_g,\bm{r}_2)   \, \bm{p}_1  = - \mbox{Im}\{ {\bm{p}_2}\cdot \bm{E}_1\} 
\end{align}
and 
\begin{align}
\label{conj_2}
    \mbox{Re}\{ {\bm{p}_1}\cdot \bm{E}_2\} =&\bm{p}_2^T \underline{\underline{G}}_e(\bm{r}_2,\bm{r}_1) \bm{p}_1
    +\bm{p}_2^T   \,   \underline{\underline{G}}_e(\bm{r}_2,\bm{r}_g)\,\mbox{Re}\{\underline{\underline{\alpha}}_{\rm eff}  \} \,\underline{\underline{G}}_e(\bm{r}_g,\bm{r}_2)   \, \bm{p}_1 
    = \mbox{Re}\{ {\bm{p}_2}\cdot \bm{E}_1\} 
\end{align}
Thus implying that  ${\bm{p}_1}\cdot \bm{E}_2 = ({\bm{p}_2}\cdot \bm{E}_1)^*$ where superscript $*$ denotes complex conjugation. Consequently, in the absence of material loss ($\Gamma=0$), 
$\mathcal{R}$ is purely imaginary. Thus, we obtain $\mathcal{R}=1$ when $\mbox{Re}\{ {\bm{p}_1}\cdot \bm{E}_2\} =0$ and $\mathcal{R}=0$ when $\mbox{Im}\{ {\bm{p}_1}\cdot \bm{E}_2\} =0$
In general, Eqs.~(\ref{firstpart_lorentz}) and (\ref{2ndpart_lorentz}) yield that
\begin{align}
\label{sub_lorentz}
    \bm{p}_1 \cdot \bm{E}_2- \bm{p}_2 \cdot \bm{E}_1 =
          {\bm{p}_2}^T   \,   \underline{\underline{G}}_e(\bm{r}_2,\bm{r}_g)\, \left[\underline{\underline{\alpha}}_{\rm eff}-\underline{\underline{\alpha}}^T_{\rm eff}  \right] \,\underline{\underline{G}}_e(\bm{r}_g,\bm{r}_2) \, \bm{p}_1
\end{align}
%
% We can write (\ref{sub_lorentz}) as:
%
% \begin{align}
% \label{Re_sub_lorentz}
%     \bm{p}_1 \cdot \bm{E}_2- \bm{p}_2 \cdot \bm{E}_1 = \frac{1}{det(\underline{\underline{\alpha}}^{-1}_{eff})} \,
%           {\bm{p_2}}^T   \,   \underline{\underline{G}}_e(\bm{r_2},\bm{r_g})\,
%           \left[Adj\left(\underline{\underline{\alpha}}^{-1}_{eff}\right)
%           -
%         Adj\left(\underline{\underline{\alpha}}^{-T}_{eff}\right) \right] \,\underline{\underline{G}}_e(\bm{r_g},\bm{r_2}) \, \bm{p_1}
% \end{align}
%
Finally,
\begin{align}
    \mathcal{R}=\frac{ {\bm{p}_2}^T   \,   \underline{\underline{G}}_e(\bm{r}_2,\bm{r}_g)\,
          \left[\underline{\underline{\alpha}}_{\rm eff}
          -
       \underline{\underline{\alpha}}^{T}_{\rm eff} \right] \,\underline{\underline{G}}_e(\bm{r}_g,\bm{r}_2) \, \bm{p}_1}{\left| {\bm{p}_2}^T   \,   \underline{\underline{G}}_e(\bm{r}_2,\bm{r}_g)\,    \left[\underline{\underline{\alpha}}_{\rm eff}
          \right] \,\underline{\underline{G}}_e(\bm{r}_g,\bm{r}_2) \, \bm{p}_1 \right|+
        \left| {\bm{p}_2}^T   \,   \underline{\underline{G}}_e(\bm{r}_2,\bm{r}_g)\,
          \left[
        \underline{\underline{\alpha}}^{T}_{\rm eff} \right] \,\underline{\underline{G}}_e(\bm{r}_g,\bm{r}_2) \, \bm{p}_1\right|}
\end{align}

\section{Electric Dyadic Green's function inside a rectangular rectangular ideally conducting cavity \label{cavity_GF} }

For a rectangular box cavity with dimensions $a\times b\times c$ along $x,y$ and $z$, respectively,  the dyadic Green's function for the electric field in the cavity is given \cite{collin1991field} as:

\begin{align}
    &\underline{\underline{\tilde{G}}}_e(\bm{r},\bm{r}') =\sum_{r=0}^\infty \sum_{s=0}^\infty \sum_{t=0}^\infty
    \frac{\bm{M}_{rst}(\bm{r})\bm{M}_{rst}(\bm{r}')}{k^2_{rst}-k^2_0} +\frac{\bm{N}_{rst}(\bm{r})\bm{N}_{rst}(\bm{r}')}{k^2_{rst}-k^2_0}-\frac{1}{k^2_0} \bm{L}_{rst}(\bm{r})\bm{L}_{rst}(\bm{r}')
\end{align}

The solenoidal modes $\bm{M_n}$ and $\bm{N_n}$ ($\bm{n}=rst$ - a triple index) modes  and rotational modes $\bm{L_n}$ are obtained from the scalar functions $\psi_{Mrst},\psi_{Nrst},\psi_{Lrst}$, which are expressed in \cite{collin1991field}. 
\begin{subequations}
\begin{align}
    &\bm{M}_{rst}=\nabla\times \bm{\hat{z}} \psi_{Mrst} \\
    & k_{rst} \bm{N}_{rst}=\nabla\times \nabla \times \bm{\hat{z}} \psi_{Nrst} \\
     & k_{rst} \bm{L}_{rst} =\nabla \psi_{Lrst}  
\end{align}
\end{subequations}
where $k_{rst}=[(r\pi/a)^2+(s\pi/b)^2+(t\pi/c)^2]^{1/2}$ and $r,s,t$ are integers.

Since the scalar functions $\psi_{Mrst},\psi_{Nrst},\psi_{Lrst}$ are real-valued, the resulting electric dyadic Green function $\underline{\underline{\tilde{G}}}_e$ is also {real-valued}.
The electric field at the field point r, inside the cavity, generated by a radiating electric dipole $\bm{p}(\bm{r}')$ located at the source point $\bm{r}'$ is
\begin{align}
    \bm{E(r)} =\omega^2 \mu_0 \underline{\underline{\tilde{G}}}_e(\bm{r,r'})  \bm{p(r')}=\underline{\underline{G}}_e(\bm{r},\bm{r}')  \bm{p}(\bm{r}')
\end{align}
where $\underline{\underline{G}}_e(\bm{r},\bm{r}')=\omega^2 \mu_0 \underline{\underline{\tilde{G}}}_e(\bm{r},\bm{r}') $

\end{document}